\documentclass[12pt]{article}
\usepackage{amsmath,amssymb,amsfonts}
\usepackage{graphicx,psfrag,epsf}
\usepackage{enumerate}
\usepackage{natbib}
\usepackage{url} 
\usepackage{nccfoots} 

\newcommand{\blind}{0}

\newcommand{\argmin}{\mathop{\mbox{argmin}}}

\newcommand{\alphah}{\hat{\alpha}}
\newcommand{\betah}{\hat{\beta}}
\newcommand{\gammah}{\hat{\gamma}}
\newcommand{\deltah}{\hat{\delta}}
\newcommand{\thetah}{\hat{\theta}}

\begin{document}

\def\spacingset#1{\renewcommand{\baselinestretch}%
{#1}\small\normalsize} \spacingset{1}


\if0\blind
{
  \title{\bf Robust Monitoring of Time Series\\
		with Application to Fraud 
		Detection\Footnote{}{Peter Rousseeuw, 
		Department of Mathematics, University 
		of Leuven, Belgium. Domenico Perrotta, Joint 
		Research Centre, Ispra, Italy. Marco Riani,
    Department of Economics, University of Parma, 
		Italy. Mia Hubert, Department of Mathematics,
		University of Leuven, Belgium.
		Peter Rousseeuw and Mia Hubert gratefully
		acknowledge the support by project C16/15/068
		of Internal Funds KU Leuven. The work of
		Domenico Perrotta was supported by the Project 
		``Automated Monitoring
		Tool on External Trade Step 5'' of the Joint
		Research Centre and the European Anti-Fraud
		Office of the European Commission, under the
		Hercule-III EU programme.
		In this paper, references to specific countries 
		and products are made only for purposes of 
		illustration and do not necessarily refer to 
		cases investigated or under investigation by 
		anti-fraud authorities.}} 
  \author{Peter Rousseeuw, Domenico Perrotta,
		Marco Riani and Mia Hubert}
  \maketitle
} \fi

\if1\blind
{
  \bigskip
  \bigskip
  \bigskip
  \begin{center}
    {\LARGE\bf Robust Monitoring of Many Time Series\\
					   	 with Application to Fraud Detection}
  \end{center}
  \medskip
} \fi

\bigskip
\begin{abstract}
Time series often contain outliers and 
level shifts or structural changes.
These unexpected events are of the utmost 
importance in fraud detection, as they may 
pinpoint suspicious transactions. 
The presence of such unusual events can easily 
mislead conventional time series analysis and 
yield erroneous conclusions.
A unified framework is provided
for detecting outliers and level shifts in
short time series that may have a seasonal 
pattern.
The approach combines ideas from the FastLTS 
algorithm for robust regression with alternating 
least squares. The double wedge plot is proposed, 
a graphical display which indicates outliers and 
potential level shifts. 
The methodology was developed to detect
potential fraud cases in time series
of imports into the European Union, and is 
illustrated on two such series.
\end{abstract}

\vfill

\spacingset{1.45} 

\section{Introduction}
\label{sect:intro}
\begin{figure}[!ht]
\begin{center}
\begin{tabular}{cc}
\noalign{\vskip0.5cm}
\includegraphics[width=0.47\textwidth]
  {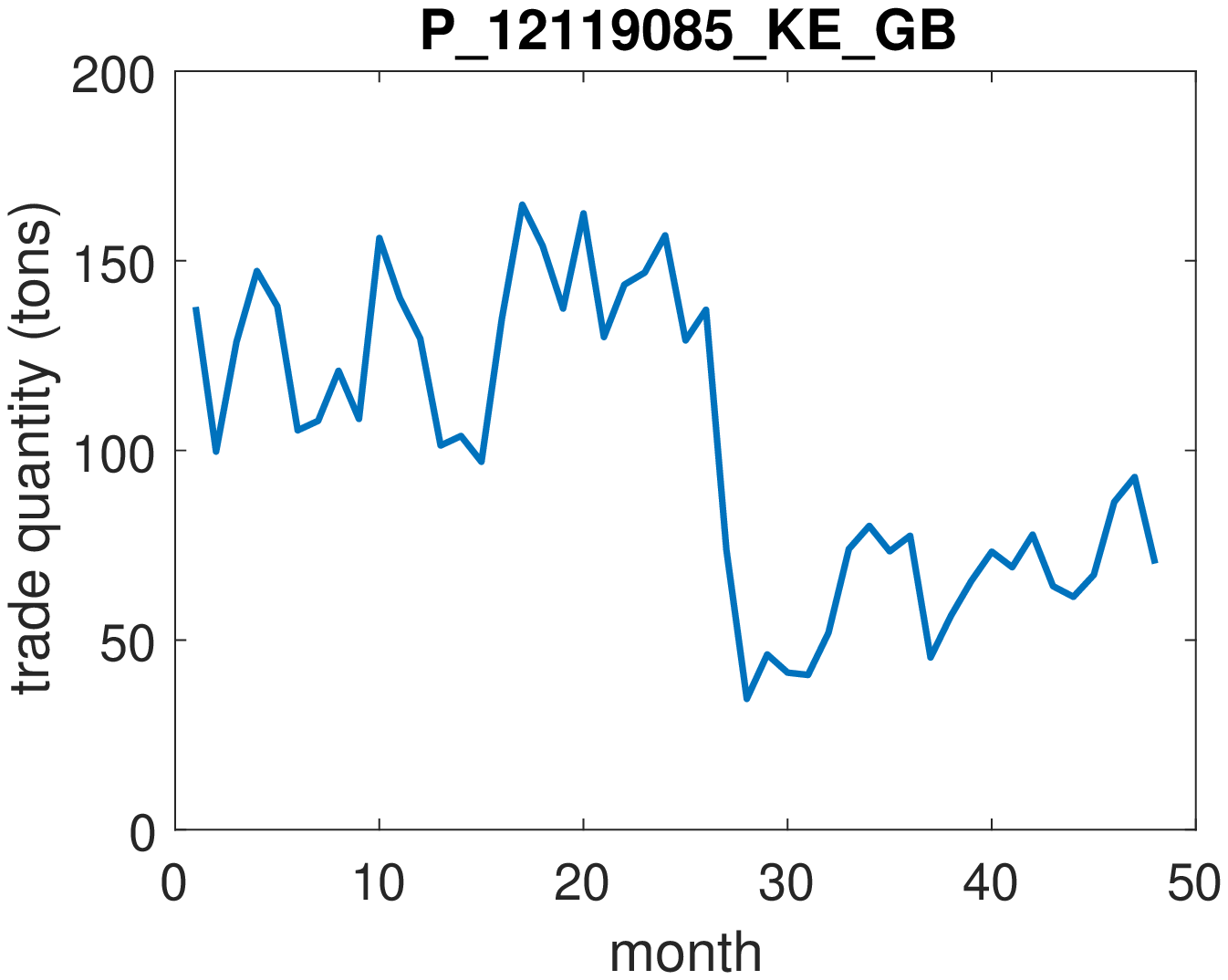} &
\includegraphics[width=0.47\textwidth]
  {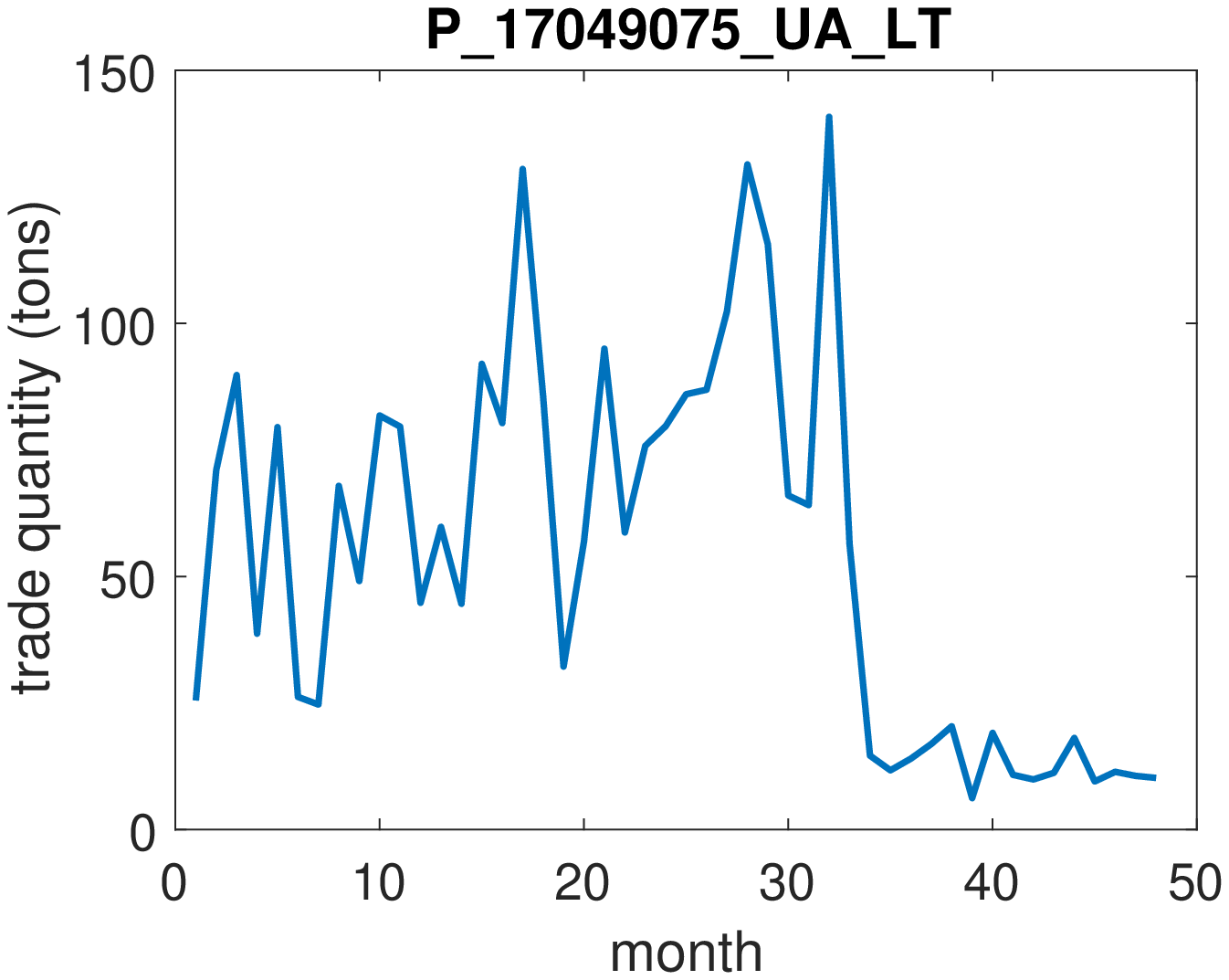}\\
(a) & (b)\\		
\noalign{\vskip-0.6cm}
\end{tabular}
\end{center}
\caption{Monthly trade volumes of two products 
imported in the European Union in a 
four-year period: 
(a) imports of plants used primarily in 
perfumery, pharmacy or for insecticidal, 
fungicidal or similar purposes, from Kenya into 
the UK (P12119085-KE-GB); (b) import of sugars 
including chemically pure lactose, maltose, 
glucose and fructose, sugar syrups, artificial 
honey and caramel, from the Ukraine into 
Lithuania (P17049075-UA-LT).}
\label{fig:trade_data}
\end{figure}

When analyzing time series one often
encounters unusual events such as outliers 
and structural changes, like those 
in Figure~\ref{fig:trade_data}.
Both series track trade volumes, and
were extracted from the official 
trade statistics in the COMEXT 
database of Eurostat. 
This database contains monthly 
trade volumes (aggregated over several 
transactions, possibly involving different 
traders) of products imported in the 
European Union (EU) in a four-year period. 
The plot titles in Figure~\ref{fig:trade_data}
specify the code of the traded product in the 
EU Combined Nomenclature classification 
(CN code), the country of origin, and the 
destination (a member state of the EU). 
The CN code determines whether the volumes are 
expressed in tons of net mass and/or other
units (liters, number of items, etc.), the 
rate of customs duty applied, and how the 
goods are treated for statistical purposes.  
The data quality is quite heterogeneous across 
countries and products, but some macroscopic 
outliers (manifest errors) have already been 
removed or corrected by statistical 
authorities and customs services.

Both of these time series exhibit a downward
level shift. Knowing when such structural 
breaks occur is important for fraud detection. 
For instance, a sudden reduction in trade 
volume may coincide with an increase for 
a related product or another country of origin, 
which could indicate a misdeclaration with the 
intent of deflecting customs duties.

There are many products and countries of 
origin in the CN classification, but not all 
of these combinations occur and the number of 
products at risk of fraud is relatively small.
Still, the number of relevant combinations of 
a product at fraud risk, a country of origin 
and a country of destination is around 16,000.
As a result, every month around 16,000 
time series need to be analyzed 
for anti-fraud purposes. 
This requires an automatic approach that
is able to report accurate information on 
outliers and the positions and amplitudes of 
level shifts, and that runs fast enough for that
time frame. The method proposed in this paper
meets those objectives, and provides a
graphical display that can be looked at whenever 
the automatic monitoring system detects a 
significant level shift. Our method follows the 
approach which first computes a robust fit to 
the majority of the data and then detects 
outliers by their large residuals, as described 
in the review paper
\citep{Rousseeuw:WIRE-Anomaly}.

A different statistical approach to monitor 
international trade data for fraud was proposed
by \citet{BaCePeCe:2016} who tested whether the 
distribution of trade volumes follows the 
Newcomb-Benford law. 
In the current paper we also take the time 
sequence of the trades into account. 
We will focus on a parametric approach to 
estimate level shifts, which differs from the 
nonparametric smoothing 
methods in \citet{Fried&Gather:2007} or robust 
methods for REGARIMA models \citep{Bianco:ARIMA}.
A popular technique is the X13 ARIMA-SEATS 
Seasonal Adjustment methodology 
\citep{Findley:1998,UScensus:X13}.
X-13 is based on automatic fitting of ARIMA models 
and includes detection of additive outliers and 
level shifts.
We will compare our results with those of X-13 
in Section \ref{sect:comparison}.
See also \citet{Galeano:Festschrift} for a review 
of robust modeling of linear and nonlinear time 
series. 

Although this paper was motivated by the need
to analyze many short time series of trade data,
we will describe the methodology more generally
so it can be applied to other types of time 
series that may be longer and can be modeled
with more parameters.

The structure of the paper is as follows. In 
Section~\ref{sect:method} we introduce our 
model and methodology for robustly analyzing a 
time series which contains a trend, a
seasonal component and possibly a level shift
in an unknown position, as well as isolated or 
consecutive outliers.
In Section~\ref{sect:airline} we illustrate the 
proposed approach using the well-known airline 
data \citep{BoxJenkins:1976}, as well as 
contaminated versions of it in order to test 
the ability of the method to detect
anomalies. In this section we also 
introduce the double wedge plot, which 
visualizes the presence of a level shift 
and outliers.
In Section~\ref{sect:tradedata} we apply our 
methodology to the time series in 
Figure~\ref{fig:trade_data}.
Section \ref{sect:comparison} compares our 
results to those obtained by a 
nonparametric method and to X-13.
The case where more than one level shift 
occurs is discussed in 
Section \ref{sect:levelshifts}.
Section~\ref{sect:conclusions} concludes, and
the Appendix proves a result about our
algorithm.

\section{Methodology}
\label{sect:method}
\subsection{The model}
\label{sect:model}
The time series $y(t)= y_t$ 
(for $t=1,\dots,T$) 
we will consider may contain the 
following terms:
\begin{enumerate}
\item a polynomial trend, i.e.
   $\sum_{a=0}^A \alpha_a t^a\;$;
\item a seasonal component, i.e.
\begin{equation}
   S_t = \sum_{b=1}^B \left( \beta_{b,1}
	\cos\left(\frac{2\pi b}{12} t\right)+ 
	\beta_{b,2} \sin\left(\frac{2\pi b}{12}
	t\right) \right)\;\;.
\label{eq:linear}
\end{equation}
When $B=1$ this is periodic with a 
one-year period, $B=2$ corresponds with 
a six-month period etc.
We assume the amplitude of the seasonal 
component varies over time in a polynomial 
way, i.e. $y_t \sim \left(1 + \sum_{g=1}^G 
\gamma_g t^g \right) S_t\;$;
\item a level shift in an unknown time point 
$2 \leqslant \delta_2 \leqslant T$, i.e.
$\delta_1 I(t \geqslant \delta_2)$ with 
$I(.)$ the indicator function.
\end{enumerate}

\noindent The general model is thus of the form
\begin{eqnarray} 
  y_t = \sum_{a=0}^A \alpha_a t^a & + & \left[ 
  \sum_{b=1}^B  \left( \beta_{b,1} 
	\cos\left(\frac{2\pi b}{12} t\right) + 
	\beta_{b,2}  \sin\left(\frac{2\pi b}{12}
	t\right) \right) \right] 
	\left(1 + \sum_{g=1}^G \gamma_g t^g \right)
	\nonumber  \\
  & + & \delta_1 I (t \geqslant \delta_2) +
	\varepsilon_t \; .
\label{eq:general}
\end{eqnarray}
One may assume that the irregular 
component $\varepsilon_t$ of the non-outliers
is a stationary random process with 
$E[\varepsilon_t] = 0$ and 
$\sigma^2 = Var[\varepsilon_t] < \infty$.
Let us collect all unknown parameters in a
vector $\theta = (\alpha_0,\alpha_1,\dots,
\beta_{1,1},\beta_{1,2},\dots,\gamma_1,
\gamma_2,\dots,\delta_1,\delta_2)$ of length $p$.
Then model~\eqref{eq:general} can be written as
$$y_t = f(\theta, t) + \varepsilon_t$$
with $f(\theta,t) = \sum_{a=0}^A \alpha_a t^a + 
 S_t (1 + \sum_{g=1}^G \gamma_g t^g) +
 \delta_1 I (t \geqslant \delta_2)\;$.
The model does not need to contain all of these
components, as some coefficients can be zero.

\subsection{The nonlinear LTS estimator} 
\label{sect:methodNLS}
Model~\eqref{eq:general} is nonlinear in 
the parameters $\beta_{b,1}$, 
$\beta_{b,2}$, $\gamma_g$ and $\delta_2$. 
As there may be outliers in the time series, 
we propose to estimate $\theta$ by means of 
the \textit{nonlinear least trimmed squares} 
(NLTS) estimator
\citep{Rousseeuw:LMS,Stromberg:Bnonlin,
Stromberg:Compnonlinear}:
\begin{equation}
  \hat{\theta}_{\mbox{NLTS}} =
	 \argmin_\theta
   \sum_{j=1}^h r^2_{(j)}(\theta) 
\label{eq:nlts}
\end{equation}
where $T/2 \leqslant h < T$ and 
$r^2_{(j)}(\theta)$ is the $j$-th smallest 
squared residual $(y_t-f(\theta,t))^2$.
Our default choice for $h$ is $[0.75\, T]$. 

The $\sqrt{n}$-consistency and asymptotic 
normality of NLTS were studied by
\cite{Cizek:2005,Cizek:2008}.
To compute the estimator, we propose to 
combine ideas from the FastLTS algorithm 
for robust linear regression 
\citep{Rousseeuw:FastLTS} with the 
alternating least squares (ALS) method.

We first describe how we use the 
alternating least squares procedure. 
We temporarily assume that the estimated 
shift time $\deltah_2$ is fixed, and that
we want to solve \eqref{eq:nlts} for a
subset of the $y_t$ with at least $p-1$ 
observations, at least one of which is
to the left of $\deltah_2$ and at
least one of which is equal or to the
right of $\deltah_2\,$. 
We denote the indices of the 
subset as $E \subset \{1,2,\dots,T\}$ 
with $\# E \geqslant p-1\;$, where $E$ 
must overlap with 
$\{1,\dots,\deltah_2-1\}$ as well as 
$\{\deltah_2,\dots, T\}$. 
These conditions are required to make
the parameters in \eqref{eq:general}
identifiable from the subset
$y_E = \{y_t; t \in E\}$.
We then go through the following steps:
\begin{enumerate}
\item{\bf [Initialization]} 
  Set $\gamma_g=0$ for $g=1,\dots,G$. 
  Then a part of \eqref{eq:general}
	drops out, leaving
	\begin{equation}
    y_t = \sum_{a=0}^A \alpha_a t^a +
		\sum_{b=1}^B \left( \beta_{b,1} 
		\cos\left(\frac{2\pi b}{12} t\right)
		+ \beta_{b,2} 
		\sin\left(\frac{2\pi b}{12} t\right)
		\right) + \delta_1
		I (t \geqslant \deltah_2) +
		\varepsilon_t
  \label{eq:fulllinear}
  \end{equation}
	which is linear in the parameters
	$\alpha_a\;$, $\beta_{b,1}\;$,
	$\beta_{b,2}$ and $\delta_1\;$.
	By applying linear LS to the subset
	$y_E\,$,
  we obtain the initial estimates
	$\alphah_a^{(0)}\;$,
	$\betah_{b,1}^{(0)}\;$,
	$\betah_{b,2}^{(0)}\;$
	and $\deltah_1^{(0)}\;$.
\item {\bf [Iteration]} For $k=1,2,...$
  repeat the following steps:
\begin{itemize}
\item {\bf [ALS step A]}			
  Let $S_t^{(k-1)}=\sum_{b=1}^B \left( 
	\betah_{b,1}^{(k-1)} 
	\cos\left(\frac{2\pi b}{12} t\right)+
	\betah_{b,2}^{(k-1)}
	\sin\left(\frac{2\pi b}{12} t\right)
	\right)$
	in which the coefficients 
	$\betah_{b,1}^{(k-1)}$ and
	$\betah_{b,2}^{(k-1)}$ come from the
	previous step.
	Keeping $S_t^{(k-1)}$ fixed yields the
	model
	\begin{equation}
     y_t-S_t^{(k-1)} = \sum_{a=0}^A
		\alpha_a t^a + S_t^{(k-1)} 
		\left(
		\sum_{g=1}^G \gamma_g t^g\right) +
		\delta_1 I(t \geqslant \deltah_2) +
		\varepsilon_t
  \label{eq:lin1}
  \end{equation}
	which is linear in the parameters
	$\alpha_a\;$, $\gamma_g\;$,
	and $\delta_1\;$.
	We then apply LS using only the 
	observations in the subset $y_E\,$,
	yielding the estimates 
	$\alphah_a^{(k)}\,$, $\gammah_g^{(k)}$
	and	$\deltah_1^{(k)}\;$.
\item {\bf [ALS step B]}
  Keeping the estimated coefficients
	$\alphah_a^{(k)}\,$, $\gammah_g^{(k)}$
	and	$\deltah_1^{(k)}$ from the previous
	step fixed yields the model
  \begin{multline}
    y_t-\sum_{a=0}^A 
		\hat{\alpha}_a^{(k)}t^a
		- \deltah_1^{(k)}I(t \geqslant
		\deltah_2) \\
    = \left[\sum_{b=1}^B \left(\beta_{b,1}
		\cos\left(\frac{2\pi b}{12} t\right)+
		\beta_{b,2}
		\sin\left(\frac{2\pi b}{12} t\right)
		\right) \right] \left(1+ \sum_{g=1}^G 
		\gammah_g^{(k)} t^g\right) +
		\varepsilon_t
  \label{eq:lin2}
  \end{multline}
  which is linear in the parameters
	$\beta_{b,1}$ and $\beta_{b,2}\;$.
	We then apply LS using only the 
	observations in the subset $y_E\,$,
	yielding the estimates 
  $\betah_{b,1}^{(k)}$ and
	$\betah_{b,2}^{(k)}\;$.
	Then we go back to ALS step A.
\end{itemize}
\end{enumerate}
Let $\thetah_{k}$ be the vector of
coefficients after iteration step $k$. 
We repeat the above steps until 
$||\thetah_{k} - \thetah_{k-1}||/
 ||\thetah_{k-1}||$ is below a threshold, 
or a maximal number of iterations (say 50) 
is attained.
Here $||\cdot||$ is the Euclidean norm.

In words, ALS solves the nonlinear LS
problem of fitting \eqref{eq:general} 
to the data set $y_E$ by alternating 
between the solution of two linear LS 
fits, \eqref{eq:lin1} and 
\eqref{eq:lin2}.

Our goal is to solve the nonlinear LTS 
problem \eqref{eq:nlts}. A basic tool
for linear LTS is the \textbf{C-step}
\citep{Rousseeuw:FastLTS} which we 
now generalize to the nonlinear setting.

{\bf [C-step]} 
Start from a subset 
$H^{(k)} \subset \{1,2,\dots,T\}$
to which we fit $\thetah^{(k)}$ 
obtained by applying ALS.
Then compute the residuals 
$r_t=y_t-f(\thetah^{(k)},t)$ for the 
whole time series, that is, for 
$t=1,\dots,T$ and not just for
$t \in H^{(k)}$.
Next retain the $h$ observations 
with smallest squared residuals,
yielding the new subset $H^{(k+1)}\,$. 
Then apply ALS to  $H^{(k+1)}\,$,
yielding a new fit 
$\thetah^{(k+1)}\,$. 
It is shown in the
Appendix that the new fit 
$\thetah^{(k+1)}$
is guaranteed
to have a lower objective function
than the old fit $\thetah^{(k)}\,$.
It is possible to iterate the C-step
until convergence, which will occur in 
a finite number of steps.

Using these building blocks, we now describe 
the entire algorithm to compute the NLTS 
fit to the model~\eqref{eq:general}.
Let $t_{(1)},\ldots,t_{(S)}$ be the ordered 
indices of the possible positions 
$\delta_2$ of the level shift, for example 
the set $\{u+1,\dots,T-u\}$ for 
some $u > 0$. The algorithm then consists
of the following steps.
\begin{enumerate}
\item \textbf{Loop over all $t_{(s)}$ where 
  $s=1,\ldots,S$ and do:} 
\begin{enumerate}
\item Temporarily set 
  $\deltah_2 = t_{(s)}\,$.
\item Now loop over $m$ ranging from 1 to 
  the number of trial subsets $M$, and do:
  \begin{enumerate}
    \item Construct an elemental subset $E$ 
	  containing $p-1$ different observations.
	  This subset should contain the index
	  $t_{(s)}\,$, one observation $y_t$ 
	  with $t < t_{(s)}$ and $p-3$ 
	  observations drawn at random from the 
	  whole time series.
		Note that we impose that $t_{(s)}$
		belongs to $E$ because the purpose
		of step 1 is to select the
	  most suitable $\deltah_2 = t_{(s)}\,$.
		\item Run the initialization and ALS
		steps described above on $E$, keeping 
	  $\deltah_2 = t_{(s)}$ fixed. 
	  Then take two C-steps.
		[Two C-steps is enough at this stage,
		in line with the results of
		\cite{Rousseeuw:FastLTS}.]
	  If a singular solution is
	  obtained  during the computations, 
	  restart without increasing $m$.
  \end{enumerate}
	The choice of $M$ is a compromise 
	since the	expected number of 
	outlier-free subsets $E$ is proportional
	to $M$ but the computation speed is 
	inversely proportional to $M$.
  In our experiments we found that 
	$M=250$ was sufficient to obtain stable
	results.
\item Consider only the {\it nbest} 
  elemental subsets (among the $M$ that 
	were tried) that yielded the lowest 
	objective function so far. Apply C-steps 
	to them until convergence and store 
	these {\it nbest} solutions. 
	In our examples we found that setting 
	{\it nbest} to 10 worked well. 
\item If $s>1$ also start from the 
  {\it nbest} elemental sets found
	when investigating $t_{(s-1)}\,$,
	but this time setting
  $\hat{\delta}_2 = t_{(s)}\,$. 
  Apply C-steps to them until 
  convergence. 
\item Take the fit with the lowest
  objective among these $2\times nbest$ 
	candidates, and denote it by 
	$\hat{\theta}^{(s)}$.	
\item Store the corresponding scaled
 residuals
 \begin{equation}
   \tilde r_{t}(\thetah^{(s)}) = 
		\frac{r_{t}(\thetah^{(s)})}
		{\sqrt{\sum_{t=1}^h r^2_{(t)}
		(\thetah^{(s)})/h}} 
		\qquad \mbox{for} \qquad
		t=1, \ldots, T\;\;.
 \label{scaledresfordwplot}
 \end{equation}
 \end{enumerate}
\item \textbf{Retain overall best 
  solution}. 
  Among the fits $\hat{\theta}^{(s)}$ 
	for $s=1,\ldots, S$ take the one 
	with lowest objective function
  $\sum_{t=1}^h r^2_{(t)}
	 (\hat{\theta}^{(s)})$
  and denote it by 
	$\hat{\theta}^{opt}\,$.\\ 
  For estimating the scale of the
	error term we can use 
	$\sum_{t=1}^h r^2_{(t)}
	 (\hat{\theta}^{opt})\,$.
  But since this sum of squares only
	uses the $h$ most central residuals, 
	the estimate needs to be rescaled. 
	The variance $\sigma^2(h)$
	of a truncated normal 
	distribution containing the central 
	$h/T$ portion of the standard 
	normal is
  \[
    \sigma^2(h)= 1-\frac{2T}{h}
		\Phi^{-1} \left(\frac{T+h}{2T}
		\right) \phi \left\{
    \Phi^{-1}\left( \frac{T+h}{2T}
		\right) \right\}
  \]
  by equation 
	(6.5) in \citep{croux+r:1992}.
  Therefore we compute
  \begin{equation} 
	  \tilde{\sigma}^2 =
    \sum_{t=1}^h r^2_{(t)}
		(\hat{\theta}^{opt}) /
		(h \sigma^2(h))\;. 
		\label{LTSsigma}
  \end{equation}
  Note that this makes 
	$\tilde{\sigma}^2$ consistent, but
  not yet unbiased for small samples. 
	Therefore we include the 
	finite-sample correction factor
	from \cite{pison+va+will:2002} in 
	our final scale estimate 
	$\hat \sigma$.
\item \textbf{Locally improving the 
  shift position estimate}.
The previous steps have yielded an 
estimate $\deltah_2$ of the position
of the level shift, but it may be
imprecise. For instance, it may happen 
that the $h$-subset underlying
$\hat{\theta}^{opt}$ does not itself
contain the time points $\deltah_2$ 
or $\deltah_2+1\,$.
In order to improve the estimate we
check in its vicinity as follows:
\begin{itemize}
\item Take a window $W$ 
  around $\deltah_2$.
  For each $t^*$ in $W$, we replace
	$\deltah_2$ by $t^*$ while keeping
	the other coefficients from
	$\hat{\theta}^{opt}$ and the scale 
	estimate $\hat{\sigma}$. 
	Compute the residuals $r_t$ from
	these coefficients and let 
	$f(t^*) = \sum_{t \in W} 
	 \rho(r_t/\hat{\sigma})$
  with $\rho$ the Huber	function 
	\begin{displaymath}
   \rho(x)=\left\{
	 \begin{array}{ccc}x^2/2 &
	   \text{if} & |x|\leqslant b\\
     b|x|-b^2/2 & \text{if} &
		|x| > b \end{array}\right.
  \end{displaymath}
  In our simulations and the analysis
	of international trade time series
	(of the kind given in 
	Figure~\ref{fig:trade_data}) the
	best results were obtained with 
	$b$ equal to 1.5 or 2.
  In our implementation the defaults
	are $b=2$ and a window $W$ of width
	15. 
\item Our final $\deltah_2$ is the
  $t^*$ in $W$ with lowest $f(t^*)$.
  If it is different from the estimate 
	we had before, we recompute the
	scaled residuals.  
\end{itemize}
\item \textbf{Weighted step.}
  We apply the univariate
	outlier detection procedure described
	in \citep{gervini2002}
	and \citep{ALYZ:2015}
	to the $T$ scaled residuals
  $(y_t-f(\thetah^{opt},t))/
	 \hat \sigma\,$.
  By default we use the 99\% confidence 
	level. Alternatively, one could use the 
	thresholds obtained in 
	\citet{salini2015}.
\item \textbf{Final fit.}
We apply nonlinear LS to all the 
points that have not been flagged 
as outliers in the previous step, 
starting from the initial estimate 
$\hat{\theta}^{opt}$ and keeping
$\deltah_2$ fixed. For this we can
iterate ALS steps until convergence. 
The standard errors obtained in the 
last two ALS steps can be used for 
inference.
\end{enumerate}

Note that $h$ must be at least the
number of parameters $p$ in the 
model for identifiability.
When $h$ is as low as $T/2$ this 
means $T/p > 2$. However, for 
stability it is often recommended 
that $T/p > 5$, see e.g.
\citet{RROD:1987}.
The Matlab code of the algorithm 
can be downloaded from 
{\it /www.riani.it/rprh/}\;.
In the following sections we will
apply it to several data sets.

\section{Airline data and the 
        double wedge plot} 
\label{sect:airline}
The airline passenger data, given as 
Series G in \citet{BoxJenkins:1976}, 
has often been used in the time series 
analysis literature as an example of 
a nonstationary seasonal time series. 
It consists of $T=144$ monthly total
numbers of airline passengers from 
January 1949 to December 1960.
Box and Jenkins developed a 
two-coefficient time series model of 
factored form that is now known
as the airline model. 
In this section we will analyze these 
data using our method, and then 
contaminate the data in various ways 
to see how the method reacts.

\begin{figure}[!ht]
\begin{center}
\includegraphics[width=0.6\textwidth]
   {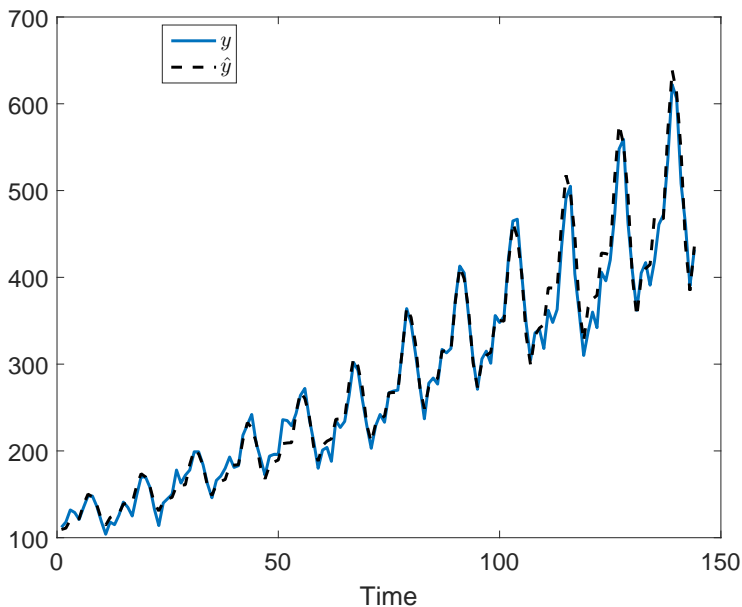}
\end{center}
\vskip-0.7cm
\caption{Airline data: observed and 
fitted values based on 
model~\eqref{eq:general} with a
quadratic trend, a quarterly seasonal 
component, and a quadratically varying 
amplitude.}
\label{fig:ADyhat}
\end{figure}

{\bf Uncontaminated data.}
We fit the data by 
model~\eqref{eq:general} with $A=2$, 
$B=4$ and $G=2$. This means that we
assume a quadratic trend, a 
quarterly seasonal component, and a 
quadratically varying amplitude.
The resulting NLTS fit
\eqref{eq:nlts}
closely follows the data, as can be 
seen in Figure~\ref{fig:ADyhat}. 
In this example no data point has 
been flagged as outlying. 
From the standard errors (not shown) 
we conclude that all coefficients 
are significant except for the 
height of the level shift.

\begin{figure}[!ht]
\begin{center}
\includegraphics[width=0.85\textwidth]
  {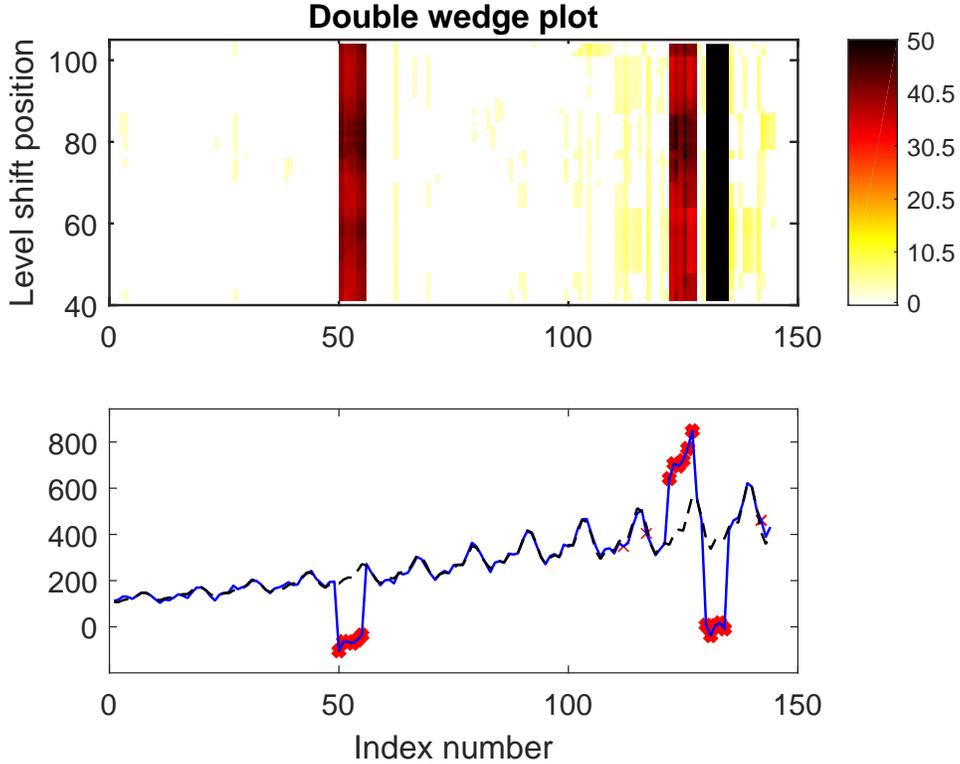}
\end{center}
\vskip-0.6cm
\caption{Airline data with contamination 1: 
double wedge plot (top) and observed and 
fitted values (bottom).}
\label{fig:ADCONT1}
\end{figure}

{\bf Contamination 1.}
We now contaminate the series by adding 
three groups of outliers,
yielding the blue curve in the bottom 
panel of Figure~\ref{fig:ADCONT1}. 
More precisely, the value 300 is 
subtracted from all responses in the 
interval $[50,55]$ while 300 is 
added on $[122,127]$ and 400 is
subtracted on  $[130,134]\,$. 
The fitted values (dotted curve) from
NLTS closely follow the observed values 
for the regular observations.
The flagged outliers are indicated by
red crosses, whose size is proportional 
to the absolute magnitude of their 
residual.
We see that all the outliers we added are 
clearly recognized as such, and they were
not used to estimate the coefficients
in the weighted step. 
Only a few regular observations received 
an absolute residual slightly above the 
cutoff value.

The top panel of 
Figure~\ref{fig:ADCONT1} is a byproduct
of the algorithm, and is useful for 
visualizing the presence of (groups of) 
outliers and a level shift.
The first step of the algorithm ranges
over all potential positions
$t_{(1)}, \ldots , t_{(S)}$ of a level
shift. These tentative positions
$t_{(s)}$ are on the vertical axis.
For any $t_{(s)}$ we plot the
absolute scaled residuals 
$|\tilde r_{t}(\thetah^{(s)})|$ given 
in \eqref{scaledresfordwplot}, in all 
of the times $t=1,2,\ldots,T$ on the 
horizontal axis.
The color in the plot depends on the 
size of that absolute residual and 
ranges from black (large residuals) over
red and yellow to white (small residuals). 
The color scale is at the right of the 
plot. Scaled residuals larger than $50$ 
are shown as if they were $50$, so that
even a very far outlier cannot affect the 
color coding. 
In the same spirit, uninformative scaled 
residuals smaller than $2.5$ are shown as
if they were 0, so in white. Of course
the user can easily modify these default
choices.

Outliers have a large absolute scaled 
residual from the robust fit, so in this 
plot isolated outliers will appear as 
dark vertical lines, and groups of
consecutive outliers as dark vertical 
bands. In this example we clearly see
the contamination.
The regular observations with scaled 
residual slightly above 2.5 do not stand
out as they are in light yellow.

{\bf Contamination 2.}
In the second contamination setting we 
introduce a persistent level shift and 
three isolated outliers, two of which
lie in the proximity of the level shift
which makes the problem harder.
For this we added the value 1300 to all
responses from $t=68$ onward, at $t=45$ 
the response is lowered by 800, at 
$t=67$ by 600, while at $t=68$ and $t=69$ 
we added an additional 800.

\begin{figure}[!ht]
\begin{center}
\includegraphics[width=0.85\textwidth]
  {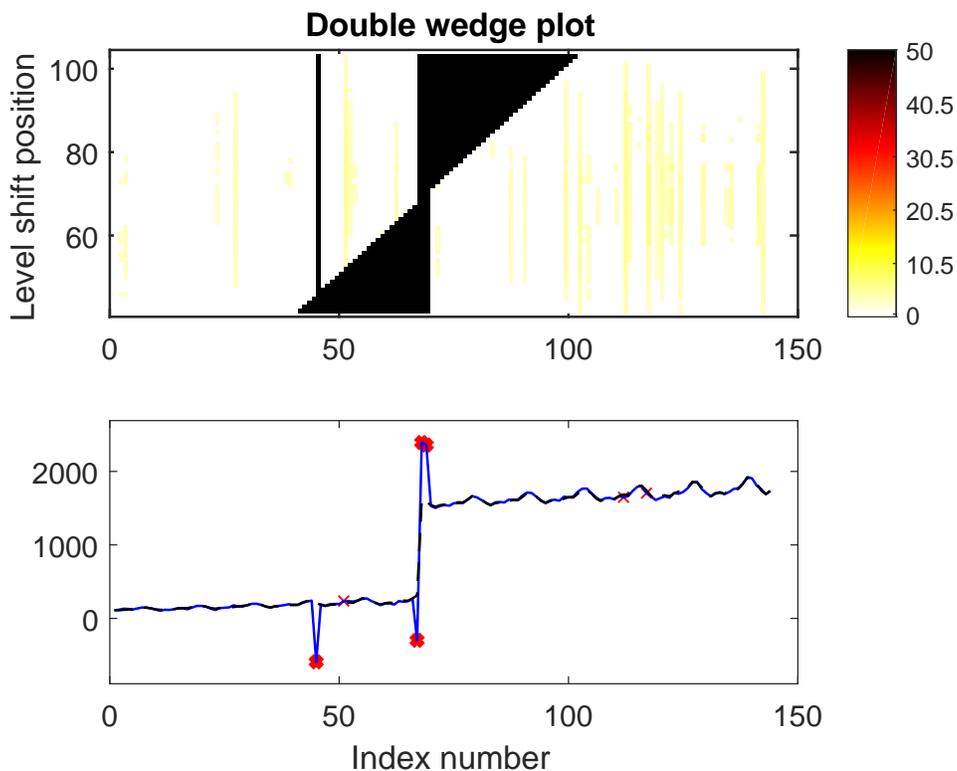}
\end{center}
\vskip-0.6cm
\caption{Airline data with contamination 2.}
\label{fig:ADCONT2}
\end{figure}

The bottom panel of 
Figure~\ref{fig:ADCONT2} shows the 
observed and fitted values.
Again all inserted outliers are clearly 
detected, and a few regular observations
have small crosses indicating that their
scaled absolute residual was slightly 
above 2.5\,. 

The plot of the absolute scaled 
residuals 
$|\tilde r_{t}(\thetah^{(s)})|$
in the top panel of 
Figure~\ref{fig:ADCONT2} now looks more
eventful with two dark triangles. 
Together these `wedges' signal a level
shift.
To understand this effect, let us
assume that the true level shift is at 
position $t^*$ and the algorithm is in
the process of checking the candidate 
$t_{(s)} = t^*-r$. 
Then the algorithm will treat the $y_t$
at $t^*-r+1,\ldots,t^*-1$ as outliers 
and the resulting robust fit (still for
that $t_{(s)}$) will show $r-1$ 
consecutive outliers.
Similarly, when the algorithm tries 
$t_{(s)} = t^*+r$ to the right of 
$t^*$, the best solutions will show 
$r$ outliers. 
As a result, when approaching the true
level shift position $t^*$ from the 
left the scaled residuals we are 
monitoring will form a dark
upward-pointing wedge, and to 
the right of the true $t^*$ we obtain
an analogous wedge pointing downward.
In the top panel of 
Figure~\ref{fig:ADCONT2}.
we observe two opposite wedges tapering 
off in the proximity of the true level 
shift position, around $t=68$.
In this region we observe a small 
rectangle (centered at position 68) 
bridging the two wedges. The rectangle 
is due to the two outliers in the 
proximity of the level shift.
The isolated outlier at position 45 
yields a single dark vertical line 
like those in 
Figure~\ref{fig:ADCONT1}.

\begin{figure}[!ht]
\begin{center}
\begin{tabular}{cc}
\includegraphics[width=0.47\textwidth]
  {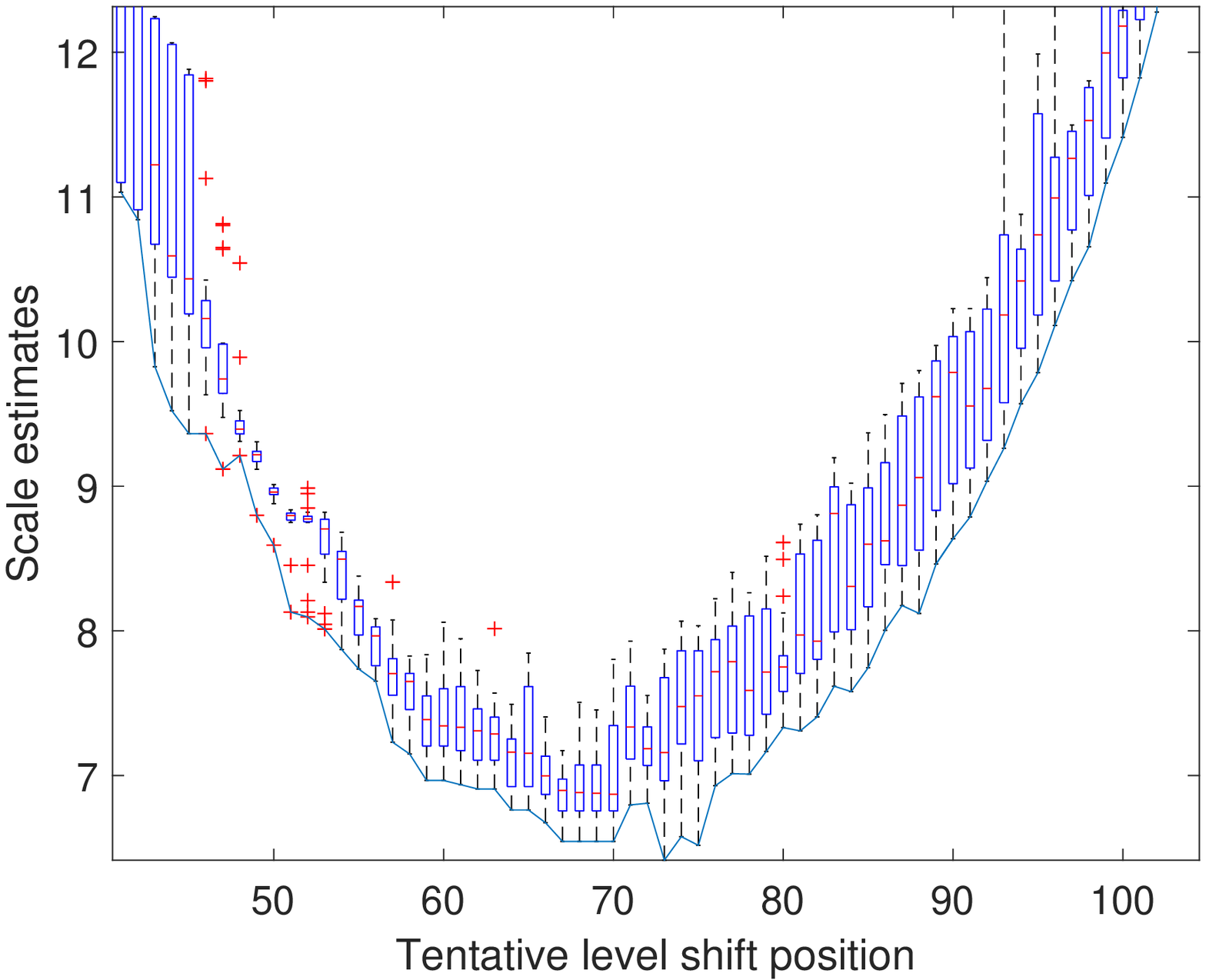} & 
\includegraphics[width=0.47\textwidth]
  {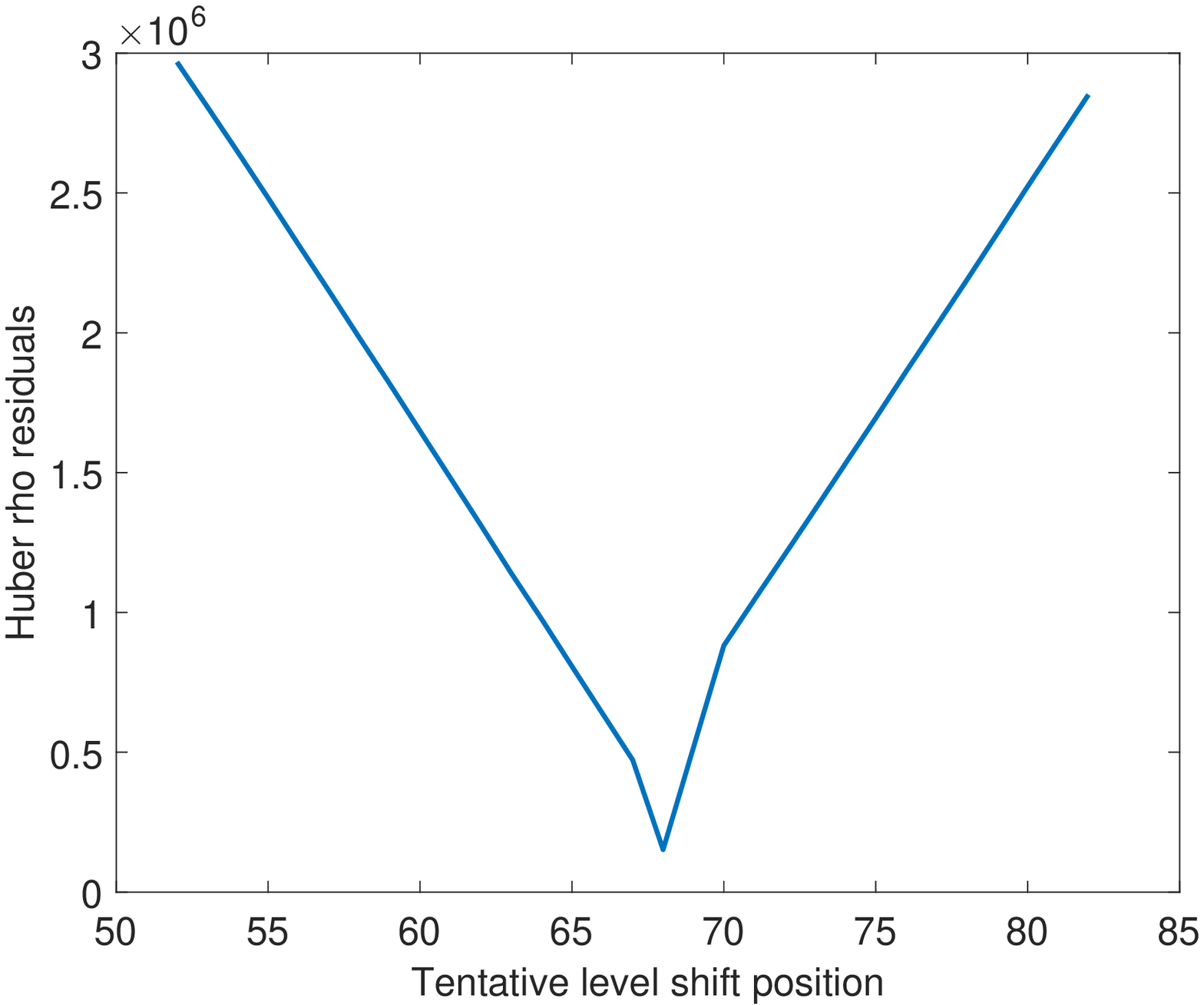} \\
(a) & (b)\\		
\noalign{\vskip-0.6cm}
\end{tabular}
\end{center}
\caption{Airline data with contamination 
2: (a) boxplots of the 20 lowest objective
function values attained at each 
$t_{(s)}\,$; 
(b) local improvement of the shift 
position estimate.}
\label{fig:ADCONT2box}
\end{figure}

Panel (a) of 
Figure~\ref{fig:ADCONT2box} shows the 
boxplots of the objective function 
$\sum_{t=1}^h r^2_{(t)}
 (\hat{\theta}_j^{(s)})$
attained by the $2\times nbest = 20$ best
solutions $\hat{\theta}_j$ in step 1(e) 
of the algorithm.
It is thus also a free byproduct of the
estimation.
If a level shift is present in the 
central part of the time series, this
plot will typically have a $U$ shape. 
In this example the lowest values of 
the trimmed sum of squared residuals 
occur in the time range 60-80. The 
continuous curve which connects the 
lowest objective value for each $s$ 
reaches its global minimum 
at $t_{(s)}=73$. 
However, the curve is quite bumpy in that
region, with several local minima and a
near-constant stretch on 67-70, so the
position of the minimum is not precise.
This kind of situation motivated the
local improvement in step 3 of the 
algorithm.
Panel (b) of Figure~\ref{fig:ADCONT2box} 
shows $f(t_{(s)}) = \sum_{t \in W} 
\rho(r_t/\hat{\sigma})$ as a function of
the tentative position $t_{(s)}$ (with
$\rho$ the Huber function with $b=2$)
on the interval 53-82.
This curve has a much better determined
minimum, in fact at $t=68$, confirming
the benefit of the local improvement 
step.

\begin{figure}[!ht]
\begin{center}
\includegraphics[width=0.85\textwidth]
  {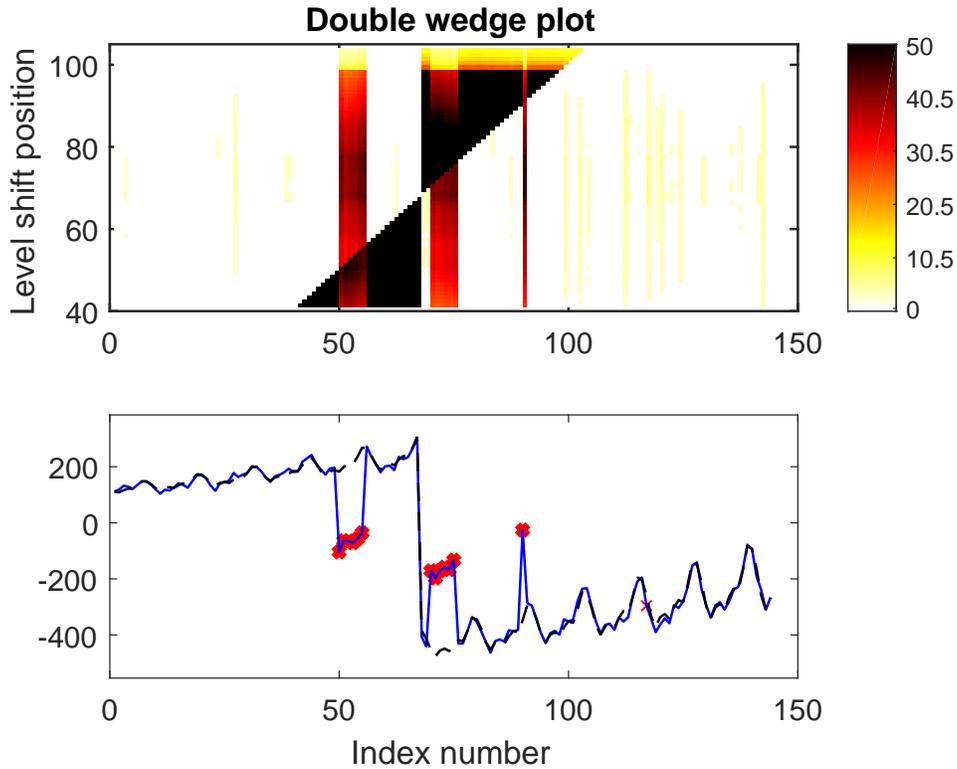}
\end{center}
\vskip-0.6cm
\caption{Airline data with contamination 3.} 
\label{fig:ADCONT3} 
\end{figure}

{\bf Contamination 3.}
In the final contaminated dataset we 
inserted a level shift and a group of 
consecutive outliers following it. 
To complicate things even more, we
also put in a stretch of contamination
to the left of the level shift, as
well as an isolated outlier.
The bottom panel of 
Figure~\ref{fig:ADCONT3} shows the 
robust fit, which succeeded in 
recovering the structure and flagging 
the outliers.
In the top panel of 
Figure~\ref{fig:ADCONT3} we see the 
typical double wedge pattern indicating
a level shift. The two reddish bands 
flag the groups of consecutive outliers, 
whereas the single line corresponds to 
the isolated outlier at $t=90$.
In this example the thick end of the 
upper wedge is yellow, so the absolute 
scaled residuals are not as large there. 
This part corresponds to a tentative
level shift of around 100, which is very 
far from the true one, and in such cases
the fit may indeed be quite different.

\section{Analysis of trade data} 
\label{sect:tradedata}
Our main goal is to analyze the many 
short time series of trade described 
in Section~\ref{sect:intro}.
After trying several model 
specifications in the class 
\eqref{eq:general} we found that the 
best results were obtained by using 
a linear trend, two harmonics, and 
one parameter to model the varying 
amplitude of the seasonal component, 
that is, $A=1$, $B=2$ and $G=1$.
Note that this yields $p=9$ parameters
including the position and height of a 
potential level shift, which is not
too many compared to the length
of the time series ($T=48$).
As an example we now apply our method 
to the time series in 
Figure \ref{fig:trade_data}.

\begin{figure}[!ht]
\begin{center}
\includegraphics[width=0.85\textwidth]
  {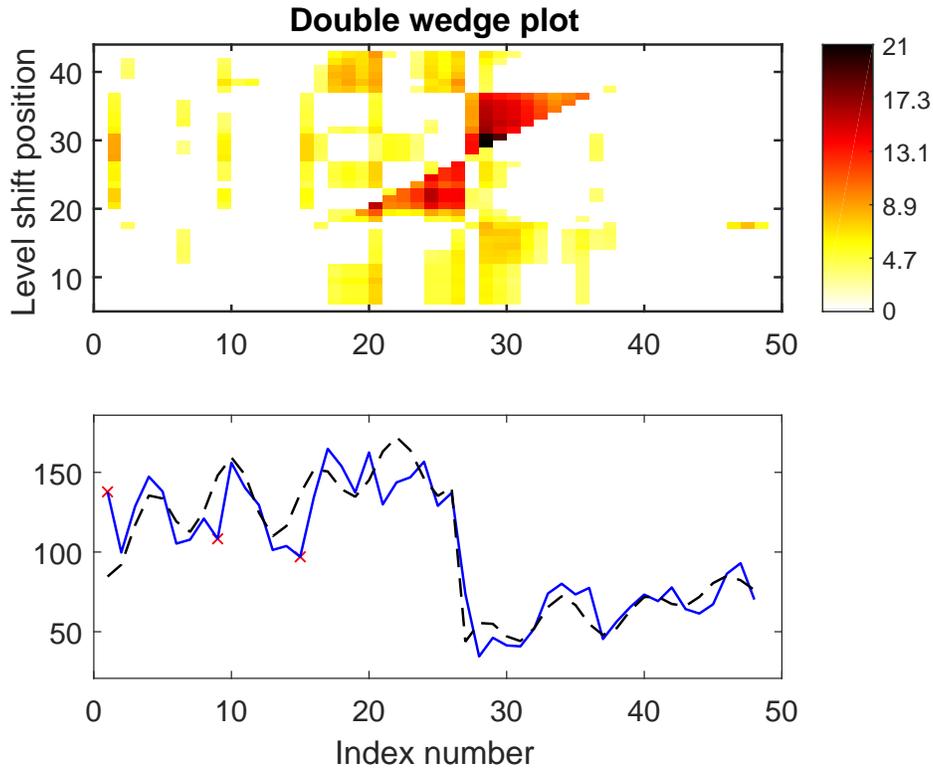}
\end{center}
\vskip-0.6cm
\caption{P12119085\_KE\_GB: double wedge
  plot (top) and observed and fitted
	values (bottom).}
\label{fig:FD1}
\end{figure}

The robust fit to series 
P12119085-KE-GB (bottom panel of 
Figure~\ref{fig:FD1}) suggests three 
moderate outliers in positions 1, 9 
and 15. The fit closely matches the
level shift which is therefore well 
captured. 
The double wedge plot in the top panel 
of Figure~\ref{fig:FD1} has two wedges 
which point to a level shift position 
around 27-28. The local refinement step
selects position $t=27$.

Columns 2--4 of Table~\ref{tab:betaest}
show the coefficients of the final fit 
together with their $t$-statistics and 
$p$-values. Most coefficients are 
significant, and in particular the 
$t$-statistic of the height of the level 
shift is quite large with $|t| = 14.7\,$.
This drop looks anomalous because
in the period considered, Kenya was the 
only country of the East African 
Community (EAC) paying high European 
import duties on flowers and related 
products including CN 12119085. On the 
other hand, Kenya is the third largest 
exporter of cut flowers in the world. 
One would therefore check for a
simultaneous  upward 
level shift in an EAC country not paying 
import duties, which could point to a
misdeclaration of origin.

\begin{table}[htbp]
\centering
\caption{Coefficient estimates, 
  $t$-statistics and $p$-values for 
	series P12119085\_KE\_GB (columns 2-4) 
	and P17049075\_UA\_LT (columns 5-7).}
\label{tab:betaest}
\vskip0.3cm
\begin{tabular}{l|rrr|rrr}
\hline
   & \multicolumn{3}{c|}{P12119085\_KE\_GB}
	 & \multicolumn{3}{c}{P17049075\_UA\_LT} \\ 
\hline
   & \multicolumn{1}{c}{Coeff}
	 & \multicolumn{1}{c}{$t$-stat}
	 & \multicolumn{1}{c|}{$p$-values}
	 & \multicolumn{1}{c}{Coeff}
	 & \multicolumn{1}{c}{$t$-stat}
	 & \multicolumn{1}{c}{$p$-values} \\
\hline	
 $\alphah_0$ & 115.27 & 25.6 & 0 & 55.14 & 14.3 & 0 \\
 $\alphah_1$ & 1.59 & 5.80 & 0 & 0.90 & 4.52 & 0 \\
 $\betah_{11}$ & -2.83 & -0.72 & 0.47 & 15.55 & 3.75 & 0.00056 \\
 $\betah_{12}$ & -12.42 & -2.65 & 0.012 & 3.61 & 0.85 & 0.40 \\
 $\betah_{21}$ & -9.07 & -1.95 & 0.059 & -32.50 & -7.64 & 0 \\
 $\betah_{22}$ & -22.60 & -4.80 & 0 & -16.06 & -3.72 & 0.00061 \\
 $\gammah_{1}$ & -0.016 & -3.72 & 0.00061 & -0.023 & -12.1 & 0 \\
 $\deltah_{1}$ & -112.62 & -14.7 & 0 & -79.41 & -13.9 & 0 \\
\hline
\end{tabular}
\end{table}

\begin{figure}[!ht]
\begin{center}
\includegraphics[width=0.85\textwidth]
  {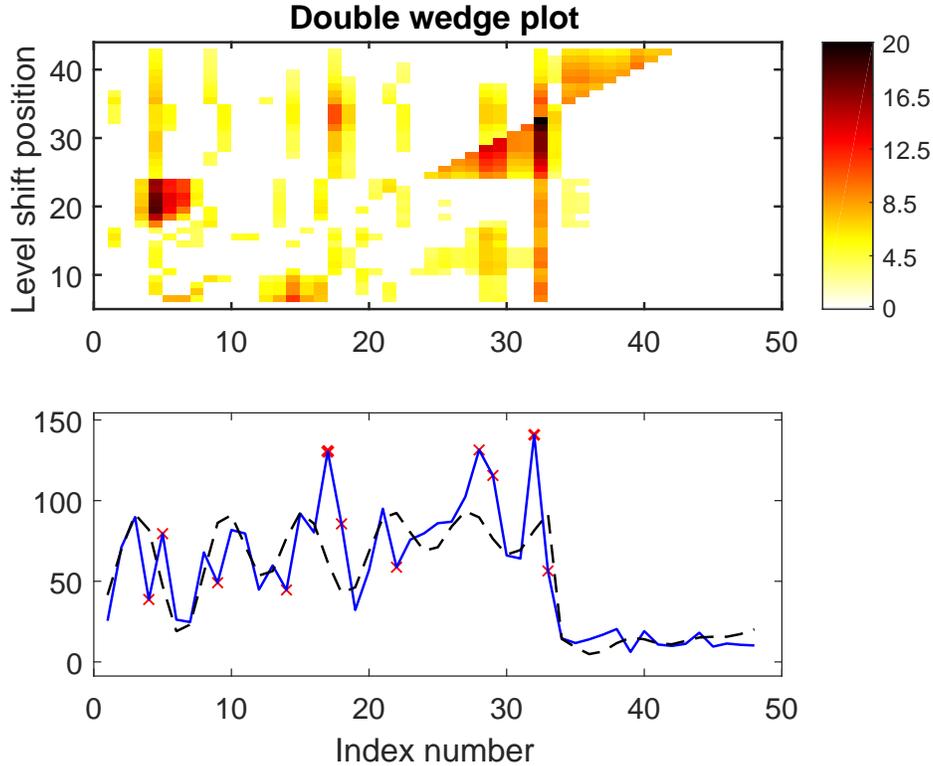}
\end{center}
\vskip-0.6cm
\caption{P17049075\_UA\_LT: double wedge
  plot (top) and observed and fitted
	values (bottom).}
\label{fig:FD2}
\end{figure}

Figure~\ref{fig:FD2} shows the results 
for the second
series, P17049075\_UA\_LT.
The double wedge plot indicates the 
presence of a level shift around 
position 35. The local refinement yields
the position $t=34$. 
Interestingly, there is a reddish line 
right before the level shift. 
This is due to an outlier in position 32 
which gets a red cross in the bottom 
panel of the figure. 
The double wedge plot also reveals a 
yellow strip at positions 29 and 30,
indicating two less extreme outliers.
Finally, the plot also shows some small 
reddish areas that
correspond to local irregularities,
for instance observations 4, 5, 17 and 
18 which are flagged as outliers in the 
bottom panel.
Columns 5--7 of Table~\ref{tab:betaest}
list the coefficients of the final fit. 
Also here most coefficients are 
strongly significant.

In this case the level shift might 
point to a different type of violation.
The market of sugar and high-sugar-content 
products, such as CN code 17049075, is 
very restricted and regulated.
The EU applies country-specific quotas 
for these products, with lower import 
duty for imports below the quota and 
a higher duty beyond this limit (tariff 
rate quotas). 
Therefore, it would be in an exporter's
interest to circumvent the quota by 
mislabeling this product as a somewhat
related product that is not under 
surveillance. In this situation one would 
check for upward level shifts in related
products from the same country.

Note that the $t$-values and $p$-values
provided by LTS can help select a model.
Table 1 fits model \eqref{eq:general}
with $A=1$ (linear trend), $B=2$ (two
harmonics), and $G=1$ (the amplitude 
varies linearly). 
If we increase $A$ we find that a quadratic
trend is insignificant, and the same for
increasing $G$.
The $t$-values indicate that there is enough 
evidence for a 6-month seasonal effect but 
are less clear on the question whether $B$ 
should be increased further for these 
short time series. 
In any case the detection of the level shift 
turns out to be stable as a function 
of $B$ here. 

\section{Comparison with other methods} 
\label{sect:comparison}
We now compare our results with those
obtained by the 
nonparametric method introduced by 
\citet{Fried:2004} and 
\citet{Fried&Gather:2007} for robust 
filtering of time series. 
For this we used the function 
\texttt{robust.filter} from the R 
package \texttt{robfilter} of 
\citet{Fried:robfilter}. The robust 
fitting methods are applied to a moving 
time window of size \texttt{width}, 
which needs to be an odd number.

\begin{table}[!ht]
\caption{P\_12119085\_KE\_GB: Positions 
of level shifts and outliers detected
by a nonparametric time series filter,
using different window widths.}
\label{tab:fried1}
\begin{center}
\begin{tabular}{cll}
\hline
Window width & Level shift position(s)
      & Outlier position(s)\\
\hline
3  & - & [ 10, 16, 17, 37, 38, 46, 47 ]\\
5  & - & [ 16,  17,  18,  47 ] \\
7  & [ 27 ] & [ 17,  18 ] \\
9  & [ 27,  37 ] & -\\
11 & [ 27 ] & -\\
\hline
\end{tabular}
\end{center}
\end{table}

\begin{table}[!ht]
\caption{P\_17049075\_UA\_LT: Positions 
of level shifts and outliers detected
by a nonparametric time series filter,
using different window widths.}
\label{tab:fried2}
\begin{center}
\begin{tabular}{cll}
\hline
Window width & Level shift position(s)
      & Outlier position(s)\\
\hline
3  & - & [ 2,  15,  44 ] \\
5  & - & [ 15,  16,  17,  28,  30,
          31,  33,  39,  40,  41 ] \\
7  & - & [ 30,  31,  33,  34 ] \\
9  & - & [ 30,  31,  33,  34,  35 ] \\
11  & [ 30 ] & [ 32 ] \\
\hline
\end{tabular}
\end{center}
\end{table}

Tables~\ref{tab:fried1} 
and~\ref{tab:fried2} report the 
position of the level shift(s) and 
outlier(s) detected with all default 
options and various choices of window
widths. We also tested different robust 
choices for the trend and scale estimation 
and some values for the \texttt{adapt} 
option which adapts the moving window 
width, with similar results.

\begin{figure}[!ht]
\vskip0.5cm
\begin{center}
\begin{tabular}{cc}
\includegraphics[width=0.47\textwidth]
  {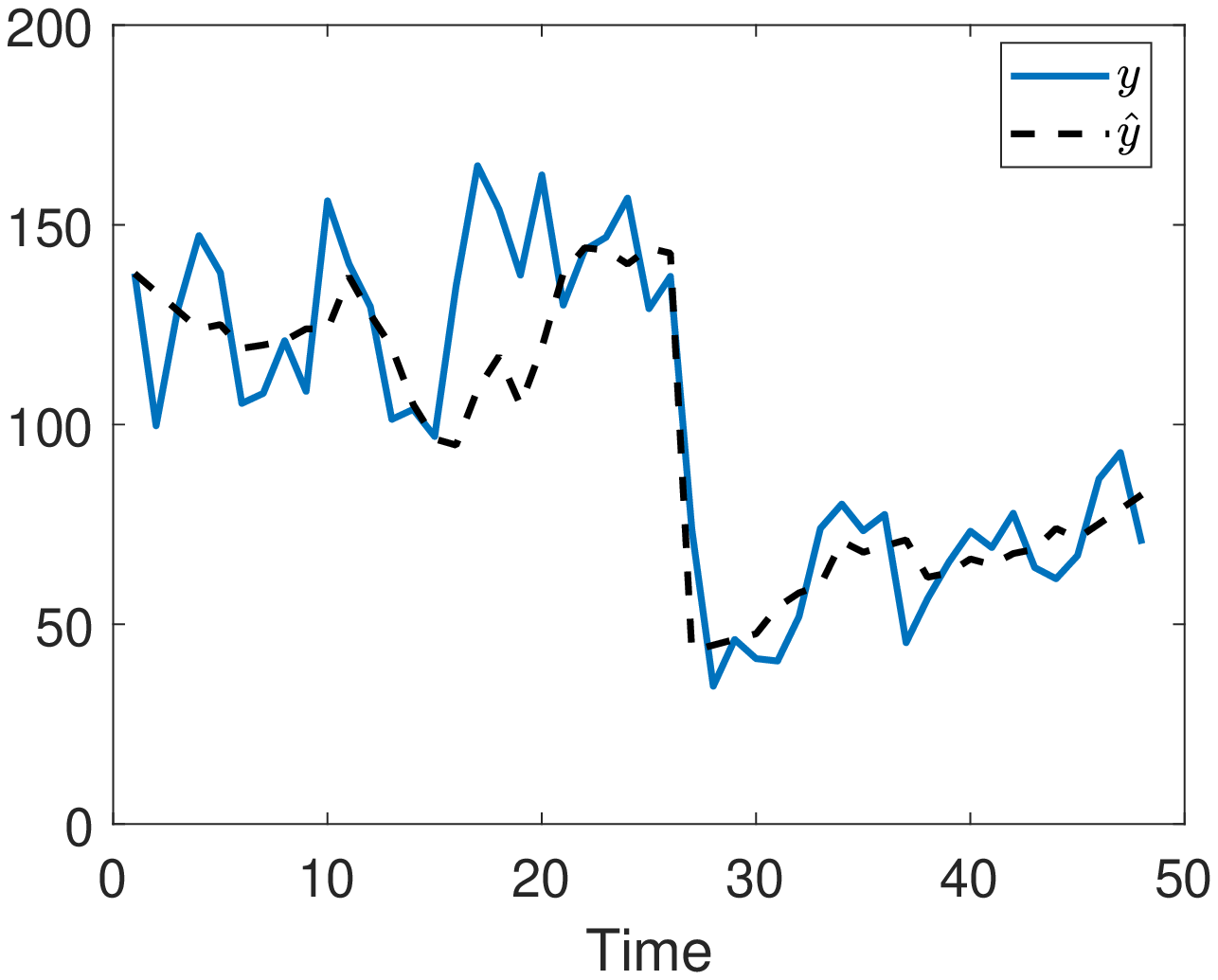}&
\includegraphics[width=0.47\textwidth]
  {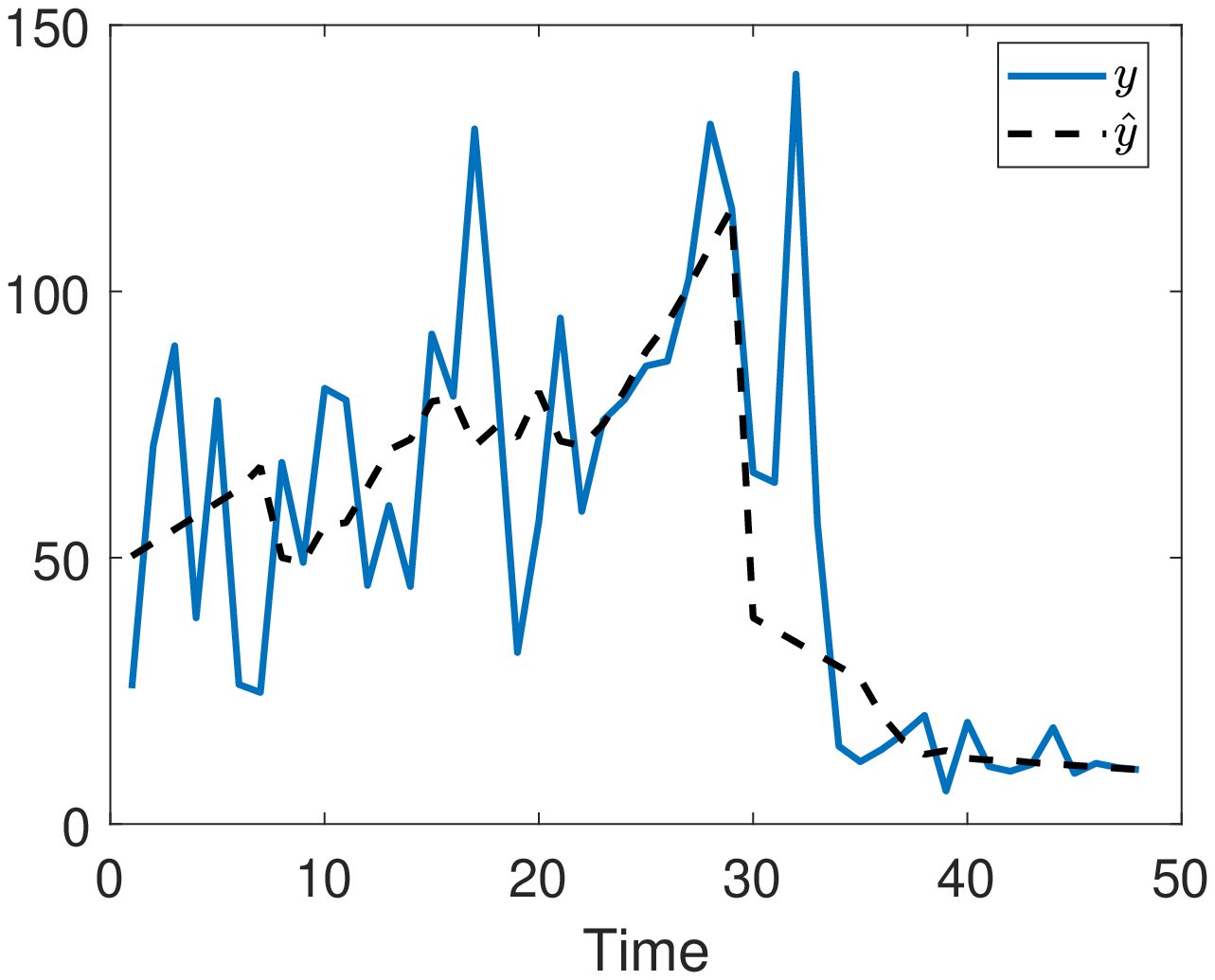}\\
 (a) & (b) \\	
\end{tabular}
\end{center}
\vskip-0.6cm
\caption{Fits obtained by a nonparametric
 time series filter for 
 (a) P12119085\_KE\_GB;
 (b) P17049075\_UA\_LT\,.}
\label{fig:Fried}
\end{figure}

Figures~\ref{fig:Fried}(a) and (b) 
show the resulting fits obtained by
the nonparametric filter, for widths
giving rise to the detection of a 
level shift (width $=7$ for 
P\_12119085\_KE\_GB and width $=11$ 
for P\_17049075\_UA\_LT).
In the first series we see that the 
level shift is detected well for the 
appropriate width, but the fit itself 
is not as tight. Also in the second
series a reasonable level shift 
position is found but the fit is not
that close to the series.
This can be explained by the fact that
a nonparametric method has no prior
knowledge about the data as it has to
work on any data set, whereas our 
parametric model benefits from 
knowledge about the typical behavior 
of trade time series.  
In that sense the comparison is not 
entirely fair.

We also run the well-known
X-13 ARIMA-SEATS method 
\citep{Findley:1998,UScensus:X13}
on both trade time series, by means 
of the R package
\texttt{seasonal} \citep{Sax:seas}
which interfaces X-13.
This method fits an ARIMA
model with a seasonal component.
In additional to the coefficients
required for the ARIMA model,
X-13 has $T$ additional 
parameters for level shifts, one
at each time point $t = 1,\ldots,T$,
plus $T$ parameters for additive
outliers (AO).
Their coefficients are estimated by
stepwise regression, so most of
them remain zero.
For detecting isolated outliers, i.e. 
outliers surrounded by 
non-outlying values, this approach
works quite well.

\begin{figure}[!ht]
\begin{center}
\vskip-0.3cm
\includegraphics[width=0.85\textwidth]
  {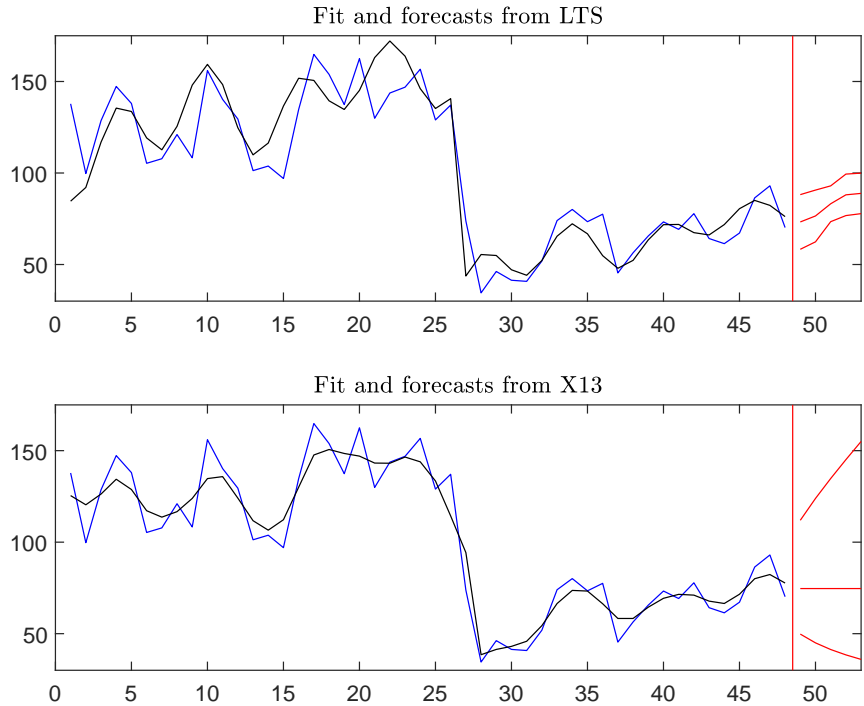}
\end{center}
\vskip-1.0cm
\caption{Trade series P12119085\_KE\_GB:
time series (blue), fit (black) and 
forecast (red) obtained by LTS (top
panel) and X-13 (bottom panel).}
\label{fig:P12fcast}
\end{figure}

\begin{figure}[!ht]
\begin{center}
\vskip-0.3cm
\includegraphics[width=0.85\textwidth]
  {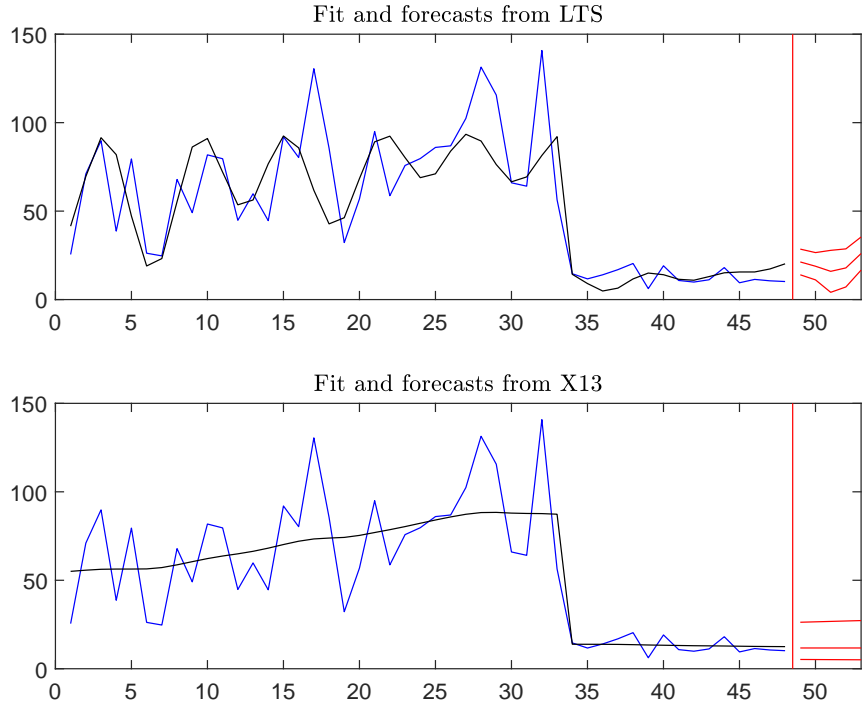}
\end{center}
\vskip-1.0cm
\caption{Trade series P17049075\_UA\_LT: 
time series (blue), fit (black) and 
forecast (red) obtained by LTS (top
panel) and X-13 (bottom panel).}
\label{fig:P17fcast}
\end{figure}

Figure \ref{fig:P12fcast} shows the
X-13 fit to the trade series 
P12119085\_KE\_GB in the lower panel,
with the LTS fit in the upper panel
for comparison.
Figure \ref{fig:P17fcast} does the 
same for P17049075\_UA\_LT.
The blue curves are the time series,
and the fits are in black. 
In both cases X-13 does detect the
level shift.
It obtains the model \mbox{(0 1 1)} 
which only has a moving average and no
seasonal component.
As a result
its forecast (shown in red) has
no seasonal component either.
Note that in Figure \ref{fig:P12fcast}   
the 90\% tolerance band around the 
forecast is much wider for X-13 than 
for LTS.

\begin{figure}[!ht]
\begin{center}
\vskip-0.3cm
\includegraphics[width=0.85\textwidth]
  {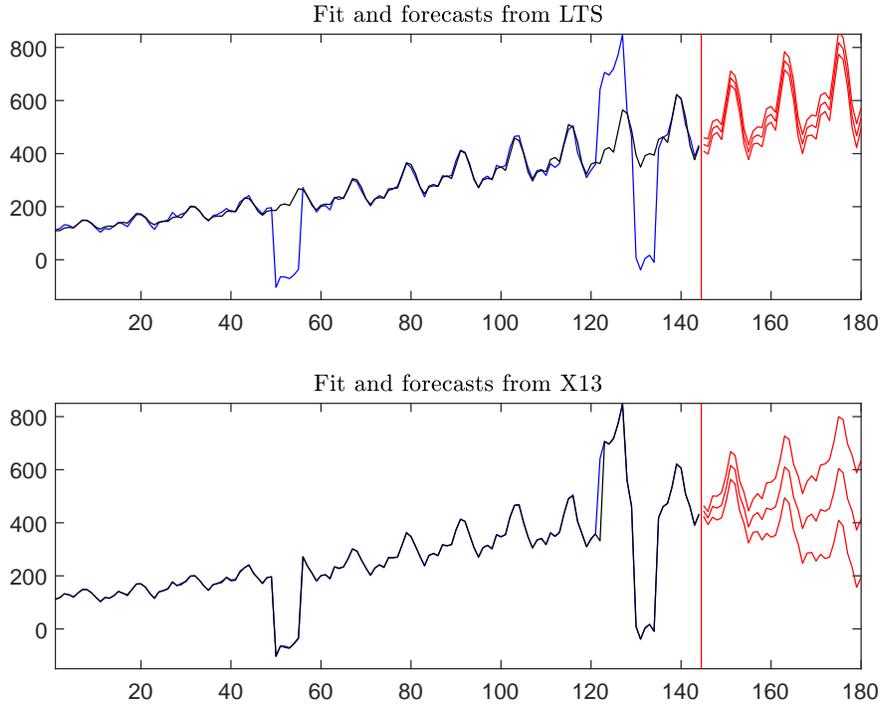}
\end{center}
\vskip-1.0cm
\caption{Airline data with contamination 1: 
time series (blue), fit (black) and 
forecast (red) obtained by LTS (top
panel) and X-13 (bottom panel).}
\label{fig:ADcont1FORE}
\end{figure}

Let us now return to the airline data
with contamination 1 described in
Section \ref{sect:airline}.
The results of LTS were
shown in Figure~\ref{fig:ADCONT1}.
We now apply X-13 to it. 
The R-code and output are available 
in Section A.1 of the Supplementary 
Material.
The model found by X-13 is ARIMA
with (1 1 0)(0 1 0) whereas for
the uncontaminated airline data it
was (0 1 1)(0 1 1).
In this example the X-13 fit has
7 nonzero coefficients describing
level shifts, and 1 nonzero
coefficient for an AO outlier.
The bottom panel of 
Figure \ref{fig:ADcont1FORE}
shows the time series and
the X-13 fit which
accommodates the outliers.
On the other hand, the LTS fit in
the top panel follows the pattern
of the majority of the data, so
the outliers 
have large residuals from it.
Also the forecasts are quite 
different: those of LTS increase and
have a narrow tolerance band, 
while those of X-13 slightly decrease
and have wide tolerance bands.
For the uncontaminated airline data
(that is, without outliers)
the forecasts and tolerance bands of
LTS and X-13 were very similar.

Note that X-13 fits each set of
consecutive outliers by a level shift 
at the start and a level shift afterward.
That description is indeed equivalent to
the consecutive outliers representation.
Our point is that accommodating the 
outliers gives a close fit to the 
observed time series, 
but as we see here it can inflate the 
forecast band.

On the airline data without outliers,
X-13 automatically log transforms the data 
before fitting it.
On the airline data with contamination it
does not, because the time series contains
at least one negative value.
(Other transformations would be possible,
but in automatic mode X-13 only considers
the log transform.)
In this example the negative values were
due to outliers, but in fact many trade time 
series in the EU database have at least one 
zero value, correctly reflecting that a 
certain product was not imported for a month, 
which will also prevent X-13 from 
transforming the data.

To investigate this issue further, we looked 
at two ways to make the contaminated airline 
data positive. The first was to add a
constant so that the minimum of the contaminated
time series becomes 1 (we also tried 5, 10 and 50).
The second was to truncate the series from below 
at 1 (or 5, 10, 50) so the downward outliers of
contamination 1 remain visible.
However, in none of these cases did X-13 carry out
a logarithmic transform, indicating that its
transformation criterion was affected by the 
outliers.

Also note that the outliers have a large
magnitude in this example.
In response to a referee request we
also provide an example with a level
shift that is smaller than the seasonal
component, in Section A.2 of the
Supplementary Material.

\section{The case of several level shifts} 
\label{sect:levelshifts}
Our basic model \eqref{eq:general} only 
covers the situation where at most one 
level shift occurs, which is a reasonable 
assumption for short time series. 
When several level shifts can
occur, we first apply our approach 
to the original time series. 
If it detects
a level shift we can modify the time
series by undoing the break, that is, 
subtract $\deltah_1$ from all $y_t$
to the right of the level shift,
after which one can search for the next
level shift, and so on.
A detailed example of this procedure is 
shown in subsection A.3
of the Supplementary Material.

\section{Conclusions and outlook}
\label{sect:conclusions}
We have introduced a new robust 
approach to model and monitor nonlinear 
time series with possible level shifts. 
A fast algorithm was developed and applied 
to several real and artificial datasets. 
We also proposed a new graphical display,
the double wedge plot, which visualizes
the possible presence of a level shift 
as well as outliers.
This graph requires no additional 
computation as it is an automatic 
by-product of the estimation. Our 
approach thus allows to automatically 
flag outlying measurements and to detect 
a level shift, which is important in
fraud detection as these may be 
indications of unauthorized transactions.
At the European Joint Research Centre,
this methodology was validated by 
comparing its results to those of visual
inspection of many trade series by
subject-matter experts.

\section*{Supplementary Material}
The supplementary material to this 
paper contains some R code and 
worked-out examples.

\section*{Appendix}
Here we prove that a C-step (as used in
the first step of the NLTS algorithm) 
can only decrease the LTS objective 
function.

Let $H^{(k)}$ be the current $h$-subset 
with its corresponding nonlinear LS 
coefficients 
$\thetah^{(k)}=(\{\alphah^{(k)}\},
\{\betah^{(k)}\},\{\gammah^{(k)}\},
\deltah_1^{(k)},\deltah_2)$ and objective 
function $L^{(k)} = \sum_{t \in H^{(k)}}
 r^2_{(t)} (\thetah^{(k)})$.

Now consider $H^{(k+1)}$, the $h$-subset 
which contains the $h$ observations with 
smallest squared residual with respect 
to $\thetah^{(k)}$. 
Then by construction
\begin{equation}
  \sum_{t \in H^{(k+1)}} r^2_{(t)} 
	(\thetah^{(k)}) \leqslant 
	\sum_{t \in H^{(k)}} r^2_{(t)} 
	(\thetah^{(k)}) = L^{(k)}\;.  
\label{eq:cstep1}
\end{equation}

The ALS step A then yields
$\thetah^{(k+0.5)}=(\{\alphah^{(k+1)}\},
\{\betah^{(k)}\},\{\gammah^{(k+1)}\},
\deltah_1^{(k+1)},\deltah_2)$.
Since it is the LS solution of the
linear model~\eqref{eq:lin1}, 
\begin{equation}
  \sum_{t \in H^{(k+1)}}
  r^2_{(t)} (\thetah^{(k+0.5)})
	\leqslant 
	\sum_{t \in H^{(k+1)}} r^2_{(t)} 
	(\thetah^{(k)})\;.
\label{eq:cstep2}
\end{equation}

Next, ALS step B yields
$\thetah^{(k+1)}=(\{\alphah^{(k+1)}\},
\{\betah^{(k+1)}\},\{\gammah^{(k+1)}\},
\deltah_1^{(k+1)},\deltah_2)$ with
\begin{equation}
  L^{(k+1)} = \sum_{t \in H^{(k+1)}}
	r^2_{(t)} (\thetah^{(k+1)})
	\leqslant 
	\sum_{t \in H^{(k+1)}}
  r^2_{(t)} (\thetah^{(k+0.5)})\;.
\label{eq:cstep3}
\end{equation}

Combining 
\eqref{eq:cstep1}-\eqref{eq:cstep3} yields
\begin{equation*}
    L^{(k+1)} \leqslant L^{(k)} 
\label{eq:cstep4}
\end{equation*}
so the new $h$-subset $H^{(k+1)}$ has an 
objective function that is less than or 
equal to that of $H^{(k)}$. 
Note that the only way to obtain
equality is if no coefficients have changed,
in which case the iteration stops.

\section*{References}
\bibliographystyle{elsarticle-harv}


\clearpage
\pagenumbering{arabic}
%
\begin{center}
{\bf SUPPLEMENTARY MATERIAL}
\end{center}
\numberwithin{equation}{section} 
\renewcommand{\theequation}
   {A.\arabic{equation}}

\section*{A.1 R-code for the airline data with 
          contamination 1}
\label{A:Rcode}

\begin{verbatim}
> library("seasonal") 
> library("forecast")
> y = AirPassengers
> y[50:55]   = y[50:55]-300
> y[122:127] = y[122:127]+300
> y[130:134] = y[130:134]-400
> out = seas(y, forecast.save = "forecasts")
> summary(out)

Coefficients:
                    Estimate Std. Error z value Pr(>|z|)    
LS1953.Feb        -294.29555    6.88904 -42.719  < 2e-16 ***
LS1953.Aug         304.98444    6.88904  44.271  < 2e-16 ***
AO1959.Feb         310.47628    7.17862  43.250  < 2e-16 ***
LS1959.Mar         329.79392   11.04369  29.863  < 2e-16 ***
LS1959.Aug        -285.79555    6.88904 -41.486  < 2e-16 ***
LS1959.Oct        -404.90220    6.88904 -58.775  < 2e-16 ***
LS1960.Mar         385.12429   14.05921  27.393  < 2e-16 ***
LS1960.Apr          47.91993   10.15210   4.720 2.36e-06 ***
AR-Nonseasonal-01   -0.30043    0.08324  -3.609 0.000307 ***
---
SEATS adj.  ARIMA: (1 1 0)(0 1 0)  Obs.: 144  Transform: none

> forec = series(out, c("forecast.forecasts", "s12"))
> forec[1:144,1] = y
> plot(forec[,1],ylim=c(-100,800),col="blue")
> fit = trend(out) + forecast::seasonal(out)
> lines(fit,col="black")
> lines(forec[,2],col="red") # forecast
> lines(forec[,3],col="red")
> lines(forec[,4],col="red")
\end{verbatim}

\section*{A.2 Effect of a small level shift}
\label{A:smallshift}

For this example the time series has length 
$T=150$ and it is generated according to 
model~\eqref{eq:general} with $A=1$, 
$B=3$, $G=1$.
In particular $\alpha_0 = 1$,
$\alpha_1 = 1$,
$\beta_{1,1},\beta_{1,2},\beta_{2,1},
 \beta_{2,2},\beta_{3,1},\beta_{3,2}]
= [20,-20,12,-12,4,-4]$
and $\gamma_1 = 8.88\;$. 
There is one level shift of height
$\delta_1 = 13,000$ at 
time $\delta_2 = 40$. 
The error term is generated with a
signal to noise ratio of 20.

\begin{figure}[!ht]
\begin{center}
\includegraphics[width=0.9\textwidth]
  {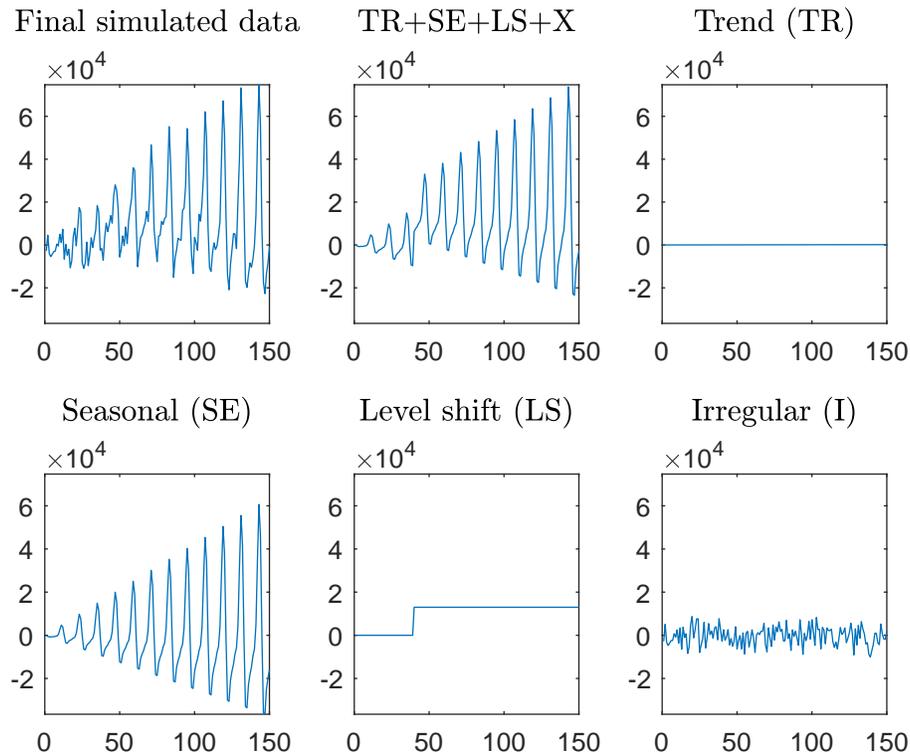}
\end{center}
\vskip-1cm
\caption{Components of the generated
data set of Section A.2.}
\label{fig:SIM6panels}
\end{figure}

All the components in 
Figure \ref{fig:SIM6panels} are shown 
using the same vertical scale so their
relative size can be seen.
The trend is increasing but appears 
horizontal if we compare it to the 
magnitude of the other components.
We see that the level shift is smaller
than the seasonal component, and similar
in size to the spread of the error term 
(called ``irregular'' in the plot).
The time series will be made available
on {\it /www.riani.it/rprh/}\;.

\begin{figure}[!ht]
\begin{center}
\vskip-0.3cm
\includegraphics[width=0.9\textwidth]
  {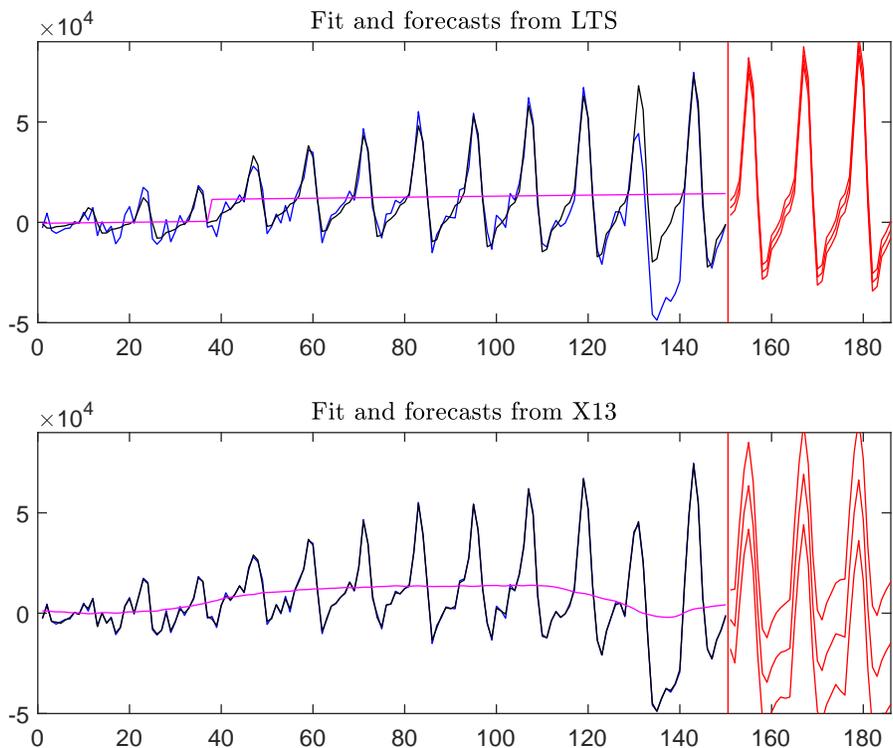}
\end{center}
\vskip-1.2cm
\caption{Data of Section A.2:
time series (blue), trend (purple),
overall fit (black), and forecast (red) 
obtained by LTS (top) and X-13 
(bottom).}
\label{fig:SIMFOREtopbottom}
\end{figure}

After generating this time series we
also create a stretch of outliers by
subtracting 29,000 from the values
at times in $[131,140]$.
The top panel of Figure 
\ref{fig:SIMFOREtopbottom} shows the 
result of the LTS procedure, with 
the time series (blue), the estimated
trend including the estimated level 
shift (purple),
the overall fit (black) and the
forecast (red). The level shift is
clearly visible, and the outliers
stand out by their sizeable residual
(look at $y_t - \hat{y_t}$ in this 
plot) for $t$ in $[131,140]$.

The bottom panel shows the X-13 fit,
which does not detect the level shift.
Instead there is a mild increase in 
the X-13 trend where the level 
shift takes place, followed by a 
mild decrease in the vicinity of
the stretch of outliers.
The X-13 forecast is stationary, 
whereas the LTS forecast has
increasing seasonal fluctuations
in line with the underlying model.
The tolerance band around the 
forecast is much wider for X-13 than 
for LTS.

\section*{A.3 More than one level shift}
\label{A:2shifts}

We now consider an example with two
level shifts.
Starting from the original airline data,
we subtract 100 from the values at 
times in $[1,30]$ and add 200 at the
times in $[100,144]$ which creates
level shifts at times 31 and 100.

\begin{figure}[!ht]
\begin{center}
\begin{tabular}{cc}
\includegraphics[height=4.5cm]
  {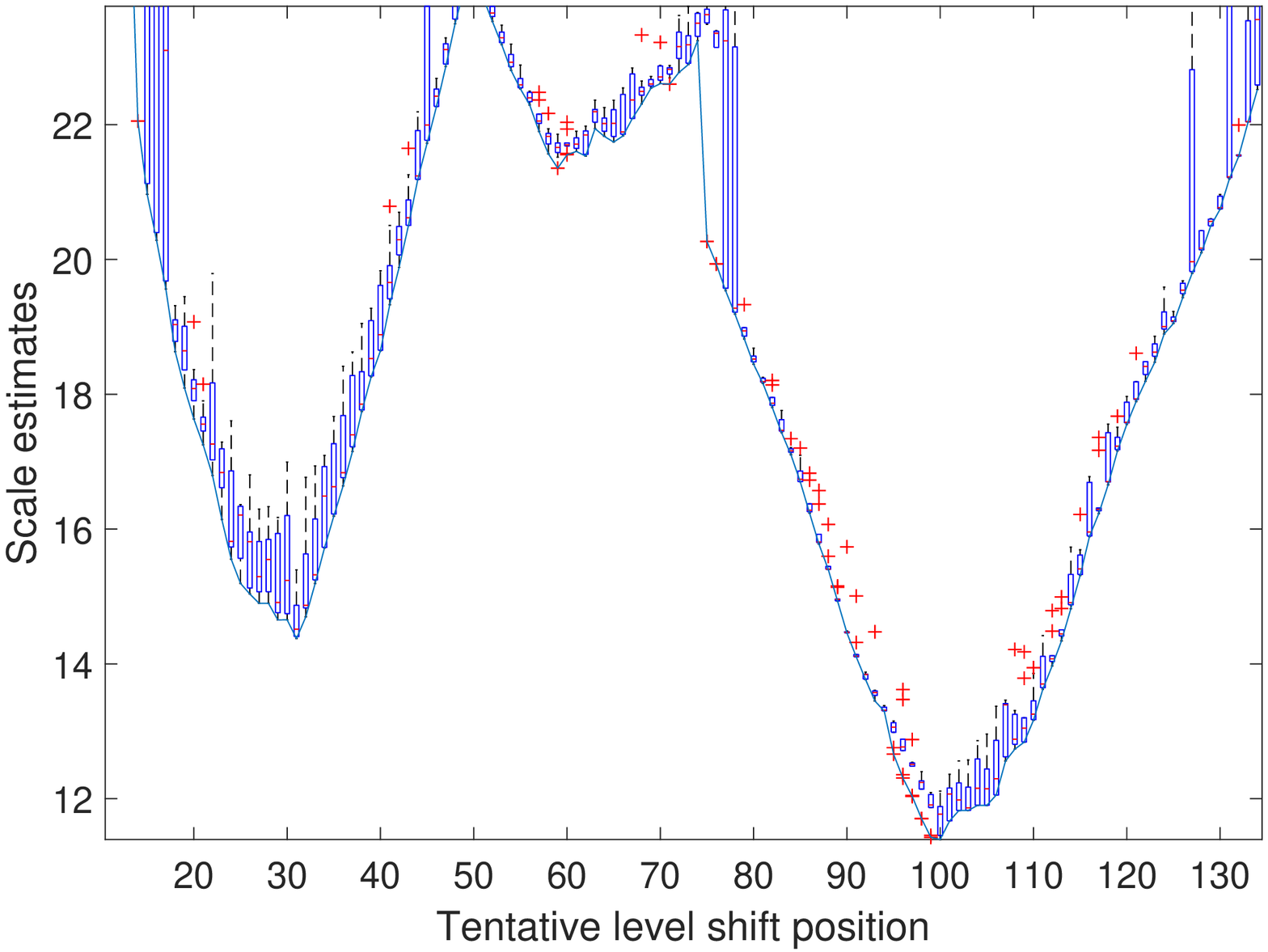} & 
\includegraphics[height=4.5cm]
  {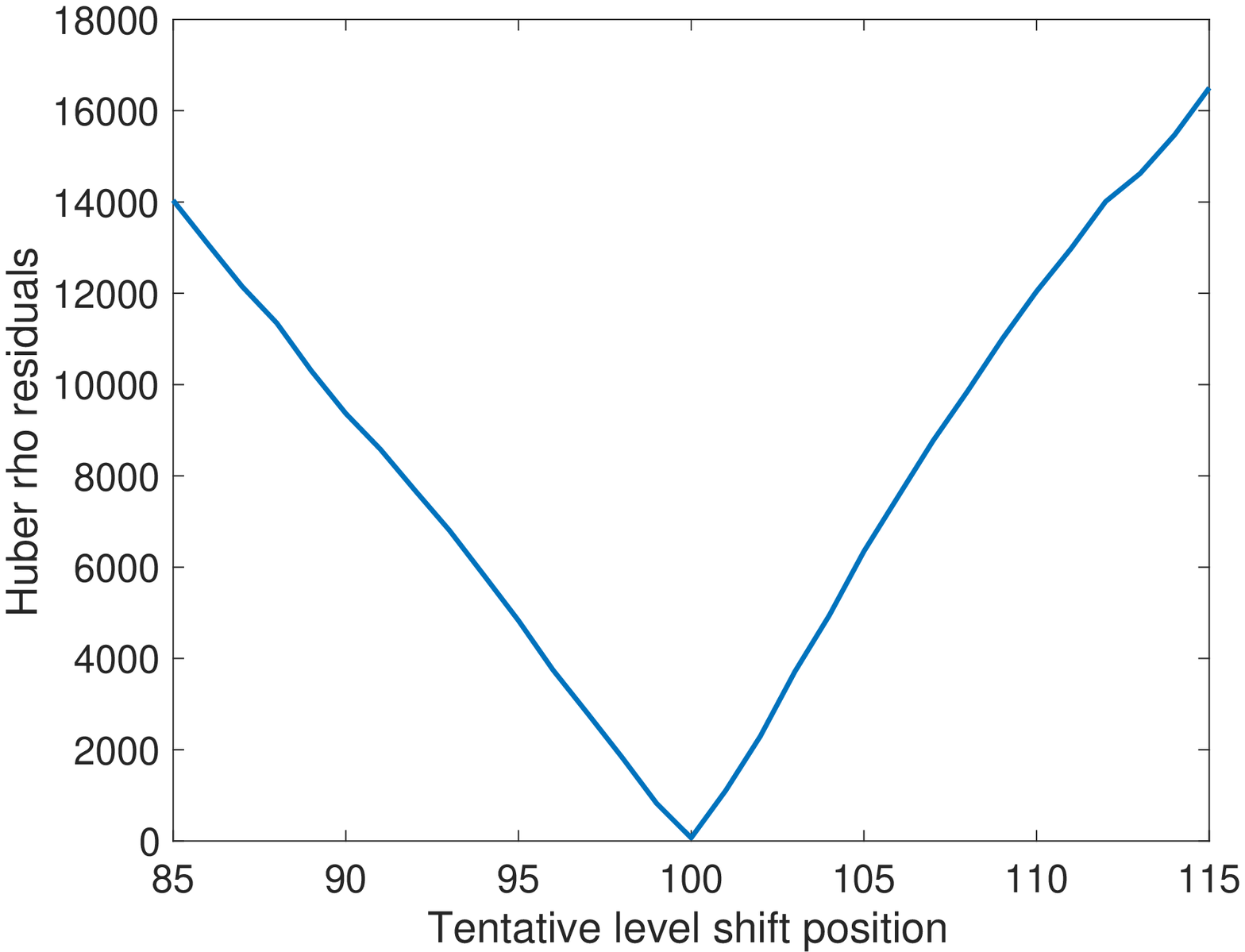} \\
(a) & (b)\\		
\noalign{\vskip-0.6cm}
\end{tabular}
\end{center}
\caption{First estimation of a level 
shift in the data of A.3:
(a) boxplots of the 20 lowest objective
function values attained at each 
$t_{(s)}\,$; 
(b) local improvement of the shift 
position estimate.}
\label{fig:ADlevelshift2}
\end{figure}

\begin{figure}[!ht]
\begin{center}
\includegraphics[width=0.85\textwidth]
  {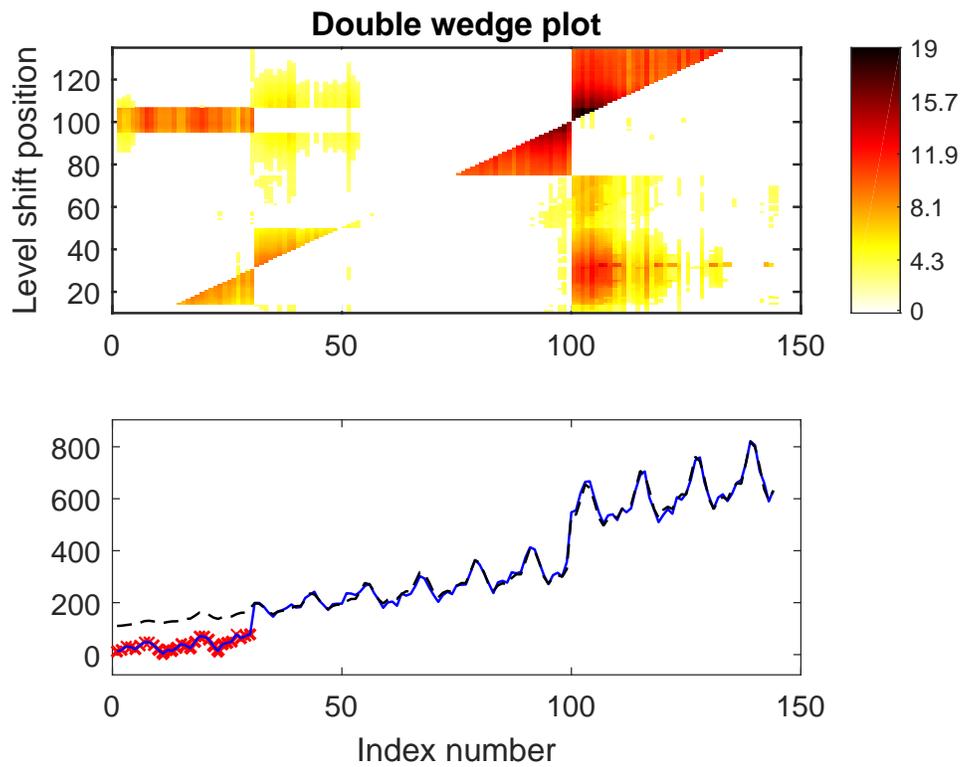}
\end{center}
\vskip-0.7cm
\caption{First fit to the data of Section
  A.3.}
\label{fig:ADlevelshift2dwplot} 
\end{figure}

Applying LTS to these contaminated
data correctly detects the level shift 
at time 100, as seen in 
Figure \ref{fig:ADlevelshift2}
with the objective function and its 
local refinement (similar to
Figure \ref{fig:ADCONT2box}).
The resulting double wedge plot
in the top panel of Figure
\ref{fig:ADlevelshift2dwplot}
actually reveals both level shifts.
Interestingly, the LTS fit in the 
lower panel of Figure
\ref{fig:ADlevelshift2dwplot}
flags the first 30 points as a stretch
of outliers.

\begin{figure}[!ht]
\begin{center}
\begin{tabular}{cc}
\includegraphics[height=4.5cm]
  {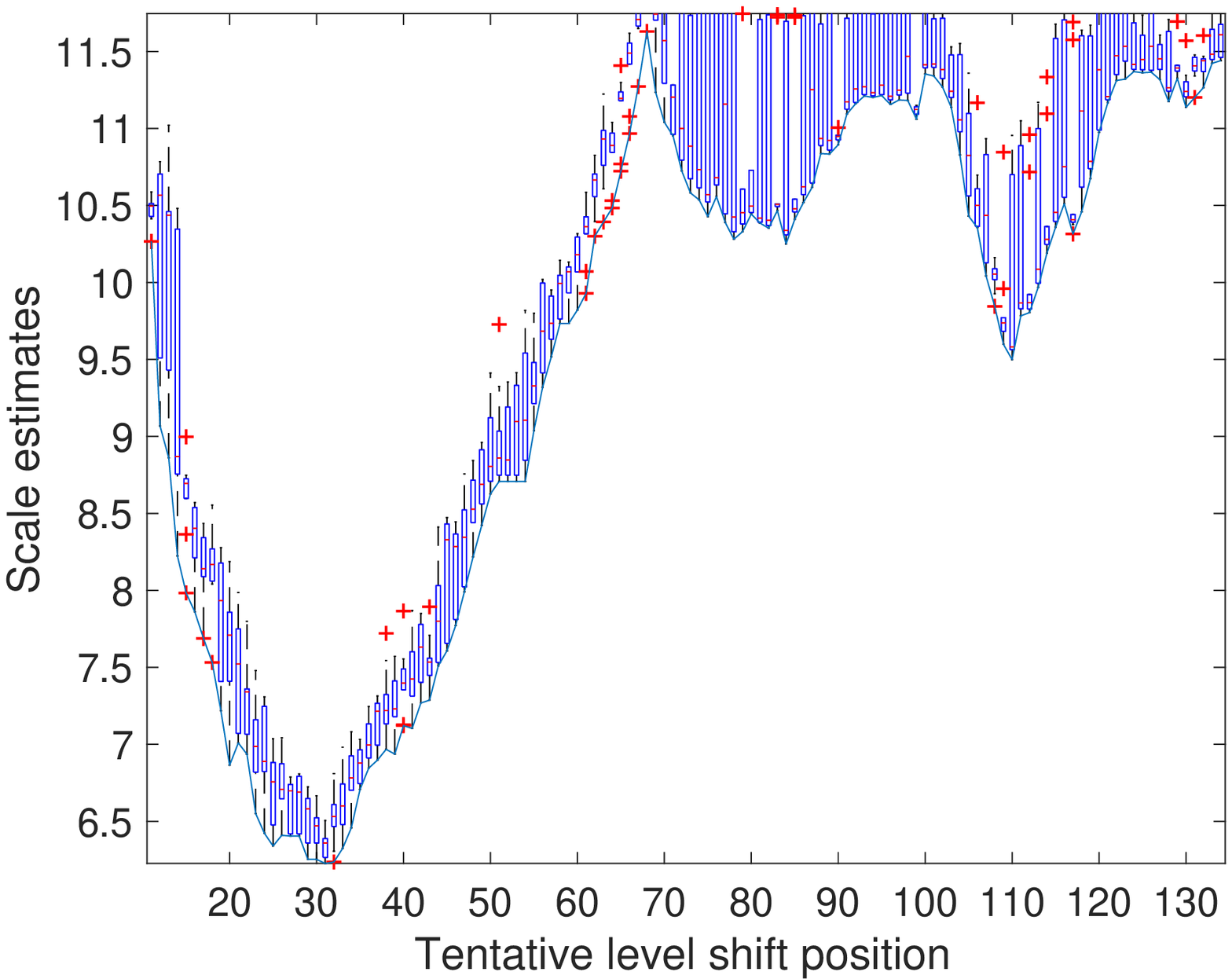} & 
\includegraphics[height=4.5cm]
  {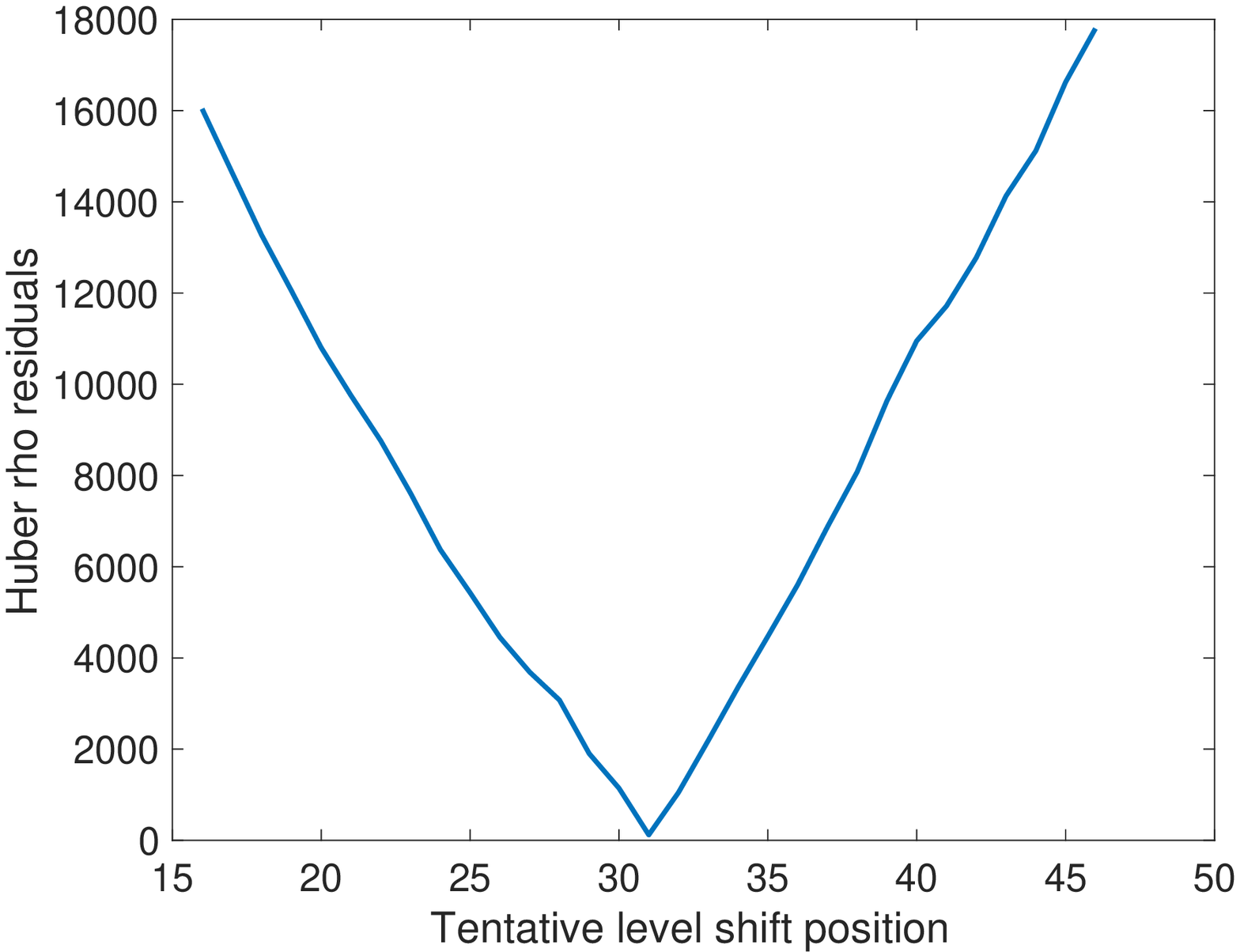} \\
(a) & (b)\\		
\noalign{\vskip-0.6cm}
\end{tabular}
\end{center}
\caption{Estimation of a level shift
in the data of A.3, after undoing the 
first level shift found.} 
\label{fig:ADlevelshift2ScaleSS}
\end{figure}

\begin{figure}[!ht]
\begin{center}
\includegraphics[width=0.85\textwidth]
  {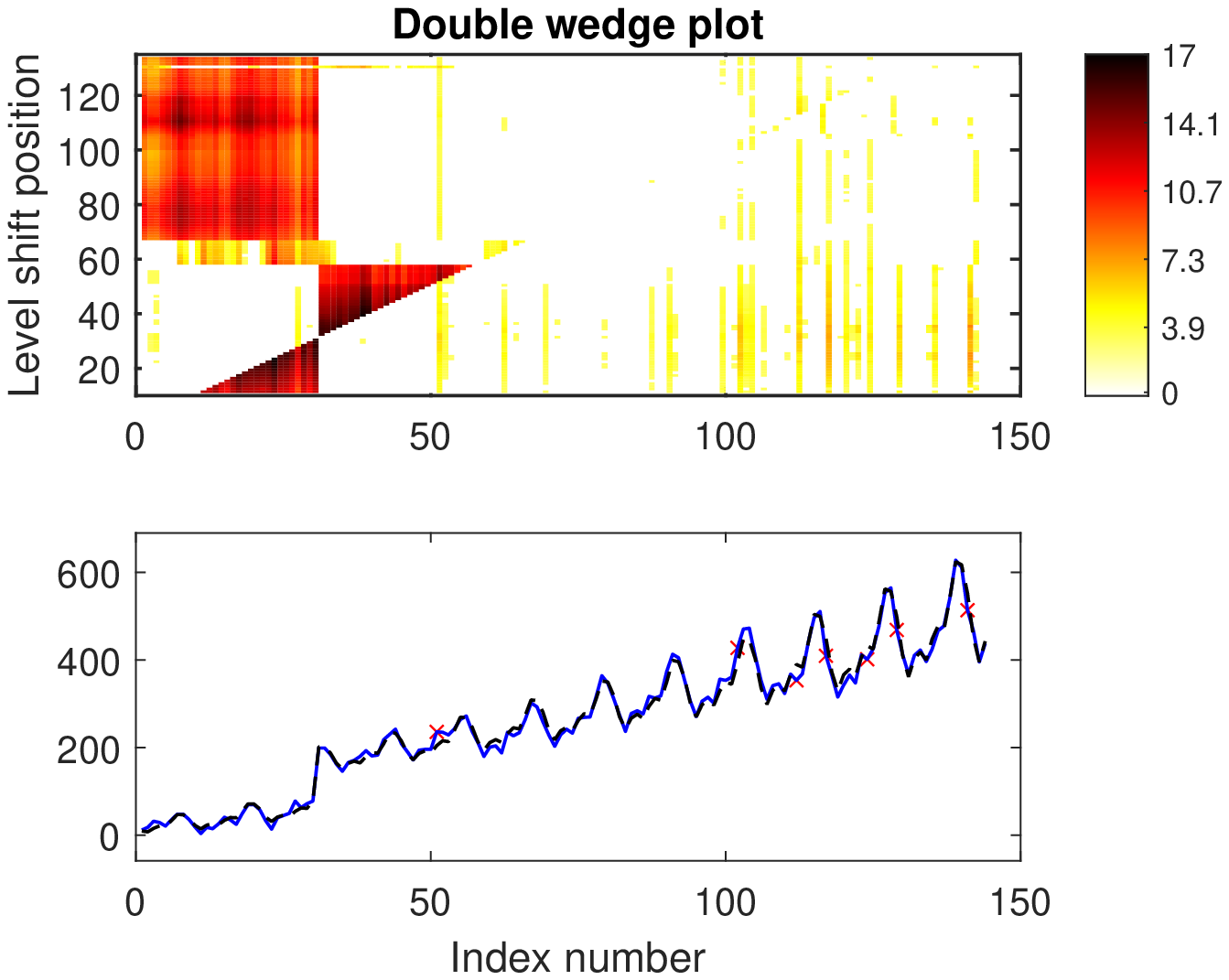}
\end{center}
\vskip-0.7cm
\caption{Fit to the data of A.3
after undoing the first estimated 
level shift.}
\label{fig:ADlevelshift2dwplotSS} 
\end{figure}

In the next step we undo the level 
shift that was found, by subtracting
$\deltah_1 = 194.47$ from $y_t$ in
all $t \geqslant \deltah_2 = 100$.
To this modified time series we again
apply LTS, which now correctly
detects the level shift at time 31
as seen in Figure 
\ref{fig:ADlevelshift2ScaleSS}.
The resulting double wedge plot in 
Figure 
\ref{fig:ADlevelshift2dwplotSS}
now shows only this level shift (since 
the other one has been removed). 
The final fit no longer shows any
outliers.
If we undo also the second level 
shift and run LTS again, no more
level shifts are found.

We also ran examples where the level
shifts had roughly the same size and 
with more than two level shifts, with 
similar results.

\end{document}